\documentclass[pdftex,twocolumn,epjc3,preprintnumbers,amsmath,amssymb]{svjour3}          

\RequirePackage[T1]{fontenc}

\smartqed  

\RequirePackage{graphicx}
\RequirePackage{mathptmx}   
\RequirePackage{flushend}
\RequirePackage[numbers,sort&compress]{natbib}
\RequirePackage[colorlinks,citecolor=blue,urlcolor=blue,linkcolor=blue]{hyperref}

\journalname{Eur. Phys. J. C}
\usepackage{amsmath}
\usepackage{amssymb}
\usepackage[normalem]{ulem}
\usepackage{bm}
\usepackage{color}
\usepackage{widetext}
\usepackage{graphicx}
\usepackage[colorlinks,citecolor=blue,urlcolor=blue,linkcolor=blue]{hyperref}
\usepackage{multirow}
\usepackage{subfigure}
\usepackage{appendix}

\newcommand{\f}{\frac}
\newcommand{\lt}{\left}
\newcommand{\n}{\nonumber}
\newcommand{\p}{\partial}
\newcommand{\rd}{{\rm d}}
\newcommand{\rt}{\right}

\begin{document}
\title{The Hawking--Page phase transitions in the extended phase space in the Gauss--Bonnet gravity} 


\author{Bing-Yu Su\thanksref{addr1}
        \and
        Yuan-Yuan Wang\thanksref{addr1}
        \and
        Nan Li\thanksref{e3,addr1}}

\thankstext{e3}{E-mail: linan@mail.neu.edu.cn}

\institute{Department of Physics, College of Sciences, Northeastern University, Shenyang 110819, China\label{addr1}}
\date{Received: date / Accepted: date}

\maketitle

\begin{abstract}
In this paper, the Hawking--Page phase transitions between the black holes and thermal anti-de Sitter (AdS) space are studied with the Gauss--Bonnet term in the extended phase space, in which the varying cosmological constant plays the role of an effective thermodynamic pressure. The Gauss--Bonnet term exhibits its effects via introducing the corrections to the black hole entropy and Gibbs free energy. The global phase structures, especially the phase transition temperature $T_{\rm HP}$ and the Gibbs free energy $G$, are systematically investigated, first for the Schwarzschild--AdS black holes and then for the charged and rotating AdS black holes in the grand canonical ensembles, with both analytical and numerical methods. It is found that there are terminal points in the coexistence lines, and $T_{\rm HP}$ decreases at large electric potentials and angular velocities and also decreases with the Gauss--Bonnet coupling constant $\alpha$.

\end{abstract}

\section{Introduction} \label{sec:intro}

Black hole thermodynamics is one of the most profound branches in modern physics, which indicates that a black hole is not simply a mathematical singularity, but should be regarded as a complicated physical system with temperature and entropy \cite{Bekenstein}. It shows deep relationship among thermodynamics, classical gravity, and quantum mechanics, and thus paves the way to our final understanding of quantum gravity \cite{law}.

However, in the first law of black hole thermodynamics, the lack of the $p$--$V$ term makes it still somehow different from traditional thermodynamics. Introducing an effective pressure is equivalent to adding a new dimension in the thermodynamic phase space, so such a theory is usually named as ``black hole thermodynamics in the extended phase space'' \cite{Cald, Kastor, Dolan1, Dolan2, Cvetic, KM, Gunasekaran}. In this framework, black hole thermodynamics is studied in the asymptotic anti-de Sitter (AdS) space with a negative cosmological constant $\Lambda$. If $\Lambda$ is allowed to change, it plays the role of a positive varying thermodynamic pressure instead of a fixed background,
\begin{align}
p=-\f{\Lambda}{8\pi}=\f{3}{8\pi l^2}, \label{P1}
\end{align}
where $l$ is the curvature radius of the AdS space, and the conjugate variable of $p$ can be effectively defined as the black hole thermodynamic volume $V$. By this means, the missing $p$--$V$ term appears in black hole thermodynamics, but turns out to be $V\,\rd p$, not the usual work term $-p\,\rd V$. Therefore, the black hole mass should be identified as its enthalpy rather than internal energy.

In the extended phase space, many remarkable similarities between black holes and non-ideal fluids were discovered, e.g., phase transitions, critical exponents, and equations of corresponding states \cite{KM, Gunasekaran}. These interesting observations aroused a large number of successive works, and almost all aspects of black hole physics were reinspected, such as the van der Waals black hole \cite{Rajagopal:2014ewa}, super-entropic black hole \cite{Hennigar:2014cfa}, superfluid black hole \cite{lambda}, reentrant phase transition \cite{trans}, heat engine \cite{Hennigar:2017apu}, throttling process \cite{li}, reverse isoperimetric inequality \cite{Dolan:2013ft}, microscopic structure \cite{Wei:2015iwa}, holographic entanglement entropy \cite{Caceres:2015vsa}, and phase transitions with non-trivial asymptotic symmetries \cite{Pedraza:2018eey} (see Ref. \cite{rev3} for reviews of recent progresses and the references therein). Currently, the research topics are mainly focusing on the black hole thermodynamics in various modified gravity theories \cite{Chen:2013ce, Zhao:2013oza, Poshteh:2013pba, Zou:2013owa, Xu:2014tja, Hennigar:2015esa, Xu:2015rfa, Li:2018rpk, Tzikas:2018cvs}.

One of the most promising modified gravity theories is the Gauss--Bonnet (GB) gravity (also referred to as the Einstein--GB gravity), which offers the leading order correction to the Einstein gravity. The GB term ${\cal G}$ is exactly the second order term in the Lagrangian of the most general Lovelock gravity. Therefore, although ${\cal G}$ itself is quadratic in curvature tensors, the equations of gravitational fields are still of second order and naturally avoid ghosts. The GB gravity possesses many important physical properties and has been heavily studied in gravitation \cite{20, 28, 45, 46, Hendi:2015pda, Khimphun:2016gsn, Konoplya:2017ymp, Doneva:2017bvd, Antoniou:2017acq} and cosmology \cite{Lidsey:2003sj, Nojiri:2005vv, Koivisto:2006xf, Koivisto:2006ai, Li:2007jm}, also with emphasis in the extended phase space \cite{Kastor:2010gq, Wei:2012ui, Cai:2013qga, Xu:2013zea, Zou:2014mha, Hendi:2015oqa, Miao:2018fke}.

In a four-dimensional manifold without boundary, the GB term is reduced to a topological invariant, $\int{\rm d}^4x\,\sqrt{-g}\,{\cal G}=\chi$, with $\chi$ denoting the Euler characteristic of the manifold. At this point, the GB term cannot have any dynamic effect in four dimensions, so it does not influence space-time structure, horizon area, global charges, and their conjugate potentials. Therefore, the GB term is usually studied in higher-dimensional physics. However, there is an exception. Albeit the GB term is irrelevant to dynamics, it does affect the thermodynamics of gravitational fields in four dimensions. The basic reason lies in the fact that, beyond the Bekenstein--Hawking formula \cite{Bekenstein}, black hole entropy receives a contribution from the GB term \cite{xiuzheng}. In this sense, the first law of black hole thermodynamics, the Smarr relation, and all the issues related to entropy will be modified. Consequently, black holes can exhibit much richer thermodynamic phenomena, especially in their phase transitions.

Among various black hole phase transitions, one of the most significant is the Hawking--Page (HP) phase transition originally studied between a Schwarzschild--AdS black hole and the thermal AdS space \cite{HP}. The black hole thermodynamics in the AdS space is quite different from that in the asymptotic Minkowski or de Sitter space. In the AdS space, large black holes have positive heat capacities and are thus thermodynamically stable, so they can be in equilibrium with the thermal background. Below a certain temperature, there is no black hole solution anymore, and the HP phase transition happens in the black hole--thermal AdS system. This phase transition was later widely investigated \cite{Witten, Birmingham:2002ph, Herzog:2006ra, Cai:2007wz, Nicolini:2011dp, Eune:2013qs, Adams:2014vza, Banados:2016hze, Czinner:2017tjq}, for example, in the charged AdS [i.e., Reissner--Nordstr\"{o}m--AdS (RN--AdS)] black holes \cite{Chamblin1, Chamblin2}. The relevant studies in the extended phase space can also be found in Refs. \cite{Spallucci:2013jja, Sahay:2017hlq, Mbarek:2018bau,sanren}.

The aim of this paper is to study the HP phase transitions in the GB gravity of the four-dimensional charged and rotating AdS [i.e., Kerr--Newman--AdS (KN--AdS)] black holes in the extended phase space (the Schwarzschild--AdS and RN--AdS black holes will also be carefully considered first). To our knowledge, this issue has not yet been available in the literature, and the major reasons are twofold. First, in physics, the GB gravity is seldom taken seriously in four dimensions; second, in mathematics, people always concentrate their attention to the black holes with simple spherical horizons. In the present work, we will explain how the GB term influences the HP phase transitions and show how to overcome the mathematical obstacle with the complicated non-spherical horizons. In short, we wish to present a thorough understanding of the HP phase transitions in the extended phase space in the GB gravity.

This paper is organized as follows. In Sect. \ref{sec:BHT}, we briefly list the thermodynamic properties of the KN--AdS black holes in the extended phase space and generally discuss the HP phase transition and the GB term. In Sect. \ref{sec:HPGB}, the HP phase transitions of the Schwarzschild--AdS, RN--AdS, and KN--AdS black holes without and with the GB term are systematically investigated in order. For the charged and rotating black holes, we work in the grand canonical ensemble with fixed electric potential and angular velocity. We conclude in Sect. \ref{sec:con}. In this paper, we work in the natural system of units and set $c=G_{\rm N}=\hbar=k_{\rm B}=1$.

\section{Black hole thermodynamics in the extended phase space} \label{sec:BHT}

In this section, we outline the thermodynamic properties of the KN--AdS black holes in the extended phase space and discuss the HP phase transition and the GB term in more detail.

\subsection{Thermodynamics of the KN--AdS black holes} \label{sec:KNAdS}

The KN--AdS black hole is the most general black hole solution in four-dimensional AdS space, with the action being
\begin{align}
\f{1}{16\pi}\int\rd^4x\,\sqrt{-g}\lt(R+\f{6}{l^2}-F_{\mu\nu}F^{\mu\nu}\rt), \n
\end{align}
where $R$ is Ricci scalar and $F_{\mu\nu}$ is the electromagnetic tensor. In the Boyer--Lindquist-like coordinates, the KN--AdS black hole metric reads
\begin{align}
\rd s^2&=-\f{\Delta_r}{\rho^2}\lt(\rd t-\f{a\sin^2\theta}{\Xi}\,\rd\phi\rt)^2+\f{\rho^2}{\Delta_r}\,\rd r^2 \n\\
&\quad+\f{\rho^2} {\Delta_\theta}\,\rd\theta^2+\f{\sin^2\theta\Delta_\theta}{\rho^2}\lt(a\,\rd t-\f{r^2+a^2}{\Xi}\,\rd\phi\rt)^2, \label{dugui}
\end{align}
where
\begin{align}
\rho^2&=r^2+a^2\cos^2\theta, \quad \Xi=1-\f{a^2}{l^2}, \quad \Delta_\theta=1-\f{a^2}{l^2}\cos^2\theta, \n\\
\Delta_r&=(r^2+a^2)\lt(1+\f{r^2}{l^2}\rt)-2mr+q^2, \n
\end{align}
and $m$, $q$, and $a$ character the mass $M$, charge $Q$, and angular momentum $J$ of the KN--AdS black hole,
\begin{align}
M=\f{m}{\Xi^2}, \quad Q=\f{q}{\Xi}, \quad J=aM=\f{am}{\Xi^2}. \label{MQJ}
\end{align}
The electromagnetic potential corresponding to the KN--AdS black hole metric is $A=-qr(\rd t-a\sin^2\theta\,\rd\phi)/\rho^2$. Moreover, Eq. (\ref{dugui}) is only valid for $a^2 < l^2$. In the limit $a^2 = l^2$, the metric in Eq. (\ref{dugui}) will become singular.

The event horizon radius $r_+$ can be determined as the largest root of $\Delta_r=0$, by which the black hole mass is expressed as $M=[(r_+^2+a^2)(r_+^2+l^2)+q^2l^2]/(2r_+l^2\Xi^2)$. To avoid naked singularity, $r_+$ must be positive. This condition sets the lower bound of the KN--AdS black hole mass as $2M^2>\sqrt{4 J^2+Q^4}+Q^2$.

Furthermore, the KN--AdS black hole entropy is obtained by the Bekenstein--Hawking formula as one quarter of the event horizon area $A$,
\begin{align}
S=\f{A}{4}=\f{\pi(r_+^2+a^2)}{\Xi}. \label{Salpha}
\end{align}
Solving $r_+$ from $S$ and using Eqs. (\ref{P1}) and (\ref{MQJ}), we can reexpress the KN--AdS black hole mass as a function of the thermodynamic quantities, $S$, $p$, $J$, and $Q$,
\begin{align}
M=\sqrt{\f{S}{4\pi}\lt[\lt(1+\f{\pi Q^2}{S}+\f{8pS}{3}\rt)^2+ \f{4\pi^2J^2}{S^2}\lt(1+\f{8pS}{3}\rt)\rt]}. \label{M}
\end{align}

Differentiating Eq. (\ref{M}) yields the first law of black hole thermodynamics in the extended phase space,
\begin{align}
\rd M=T\,\rd S+V\,\rd p+\Phi\,\rd Q+\Omega\,\rd J, \label{first}
\end{align}
where $T$, $V$, $\Phi$, and $\Omega$ are the Hawking temperature, thermodynamic volume, electric potential, and angular velocity of the KN--AdS black holes respectively,
\begin{align}
T&=\lt(\f{\p M}{\p S}\rt)_{p,J,Q}=\f{1}{8\pi M}\lt[\lt(1+\f{\pi Q^2}{S}+\f{8pS}{3}\rt)\right. \n\\
&\qquad\qquad\qquad\quad~~\left.\lt(1-\f{\pi Q^2}{S}+8pS\rt)-\f{4\pi^2J^2}{S^2}\rt], \label{T}\\
V&=\lt(\f{\p M}{\p p}\rt)_{S,J,Q}=\f{2S^2}{3\pi M}\lt(1+\f{\pi Q^2}{S}+\f{8pS}{3}+\f{2\pi^2J^2}{S^2}\rt), \label{V}\\
\Phi&=\lt(\f{\p M}{\p Q}\rt)_{S,p,J}=\f{Q}{2M}\lt(1+\f{\pi Q^2}{S}+\f{8pS}{3}\rt),\label{Phi}\\
\Omega&=\lt(\f{\p M}{\p J}\rt)_{S,p,Q}=\f{\pi J}{MS}\lt(1+\f{8pS}{3}\rt). \label{Omega}
\end{align}
Moreover, in Eq. (\ref{first}), the $p$--$V$ term has the form of $V\,\rd p$ but not $-p\,\rd V$, so the KN--AdS black hole mass $M$ should be essentially identified as its enthalpy instead of internal energy. Furthermore, the Smarr relation, as the integral form of Eq. (\ref{first}), can be obtained by a scaling argument, $M=2TS-2pV+\Phi Q+2\Omega J$.

\subsection{HP phase transition} \label{sec:HP}

With quantum effects taken into account, a black hole not only absorbs but also emits energy to external environment via the Hawking radiation mechanism \cite{Hawking2}. The exchange of energy will establish the thermal equilibrium at a fixed temperature between a stable black hole (with positive heat capacity) and the thermal AdS space.

On the one hand, as we have seen in Sect. \ref{sec:KNAdS}, the black hole mass should be regarded as enthalpy in the extended phase space, so the thermodynamic potential of interest turns out to be the Gibbs free energy. Due to the conservations of charge and angular momentum, it is not possible for a charged or rotating black hole to undergo the HP phase transition to the thermal AdS space, which carries no charge or angular momentum. As a result, any discussion of the HP phase transitions of the KN--AdS black holes must be carried out in a grand canonical ensemble, in which the electric potential $\Phi$ or angular velocity $\Omega$ is fixed and the charge $Q$ or angular momentum $J$ is allowed to vary. Therefore, the Gibbs free energy should be constructed as
\begin{align}
G(T,p,\Phi,\Omega)=M-TS-\Phi Q-\Omega J. \label{G}
\end{align}
On the other hand, since the gravitational potential of the AdS space increases at large distances, acting as a box of finite volume, the total energy of thermal AdS space is finite, and its Gibbs free energy is zero. Because the thermal equilibrium condition corresponds to the global minimum of the Gibbs free energy, the criterion of the HP phase transition is that the Gibbs free energy of the black hole vanishes,
\begin{align}
G=0. \label{panju}
\end{align}

The condition in Eq. (\ref{panju}) fixes the HP phase transition temperature $T_{\rm HP}$. It will be shown in Sect. \ref{sec:HPGB} that the Gibbs free energies of black holes decrease with temperature. Consequently, above $T_{\rm HP}$, the configuration of the black hole with negative Gibbs free energy is thermodynamically preferred; below $T_{\rm HP}$, the thermal AdS phase with vanishing Gibbs free energy is preferred and is thus stable against collapse to a black hole. This counterintuitive observation indicates that the thermal AdS space behaves more like a solid rather than an ordinary gas.

\subsection{GB term} \label{sec:GB}

The action of the KN--AdS black hole in the GB gravity reads
\begin{align}
\f{1}{16\pi}\int{\rm d}^4x\,\sqrt{-g}\lt(R+\f{6}{l^2}-F_{\mu\nu}F^{\mu\nu}+\alpha{\cal G}\rt), \n
\end{align}
with
\begin{align}
{\cal G}:=R_{\mu\nu\lambda\rho}R^{\mu\nu\lambda\rho}-4R_{\mu\nu}R^{\mu\nu}+R^2 \n
\end{align}
being the GB term, where $R_{\mu\nu\lambda\rho}$ is the Riemann tensor, $R_{\mu\nu}$ is the Ricci tensor, and $\alpha$ is the GB coupling constant.

With the GB term, the KN--AdS black hole entropy can be attained via an integral over the event horizon of the Ricci scalar $\tilde{R}$ of the two-dimensional induced metric \cite{xiuzheng},
\begin{align}
S=\f 14\int\rd\theta\rd\phi\,\sqrt{\tilde{h}}\big(1+2\alpha\tilde{R}\big), \label{gongshi}
\end{align}
where $\tilde{h}=(r_+^2+a^2)\sin\theta/\Xi$ is the determinant of the induced metric. A straightforward integral shows that the correction to the black hole entropy in Eq. (\ref{gongshi}) is neatly $4\pi\alpha$, independent of the horizon radius and shape. This is not surprising, as it is a natural result of the Gauss--Bonnet theorem applied to the two-dimensional horizon, so the integral of $\tilde{R}$ simply corresponds to its Euler characteristic. As a result, 
\begin{align}
S=\f{A}{4}+4\pi\alpha=\f{\pi(r_+^2+a^2)}{\Xi}+4\pi\alpha. \label{Swith}
\end{align}
Furthermore, from Eq. (\ref{G}), the Gibbs free energy of the KN--AdS black hole should also receive a correction with the GB term,
\begin{align}
G(T,p,\Phi,\Omega,\alpha)=G(T,p,\Phi,\Omega)-4\pi\alpha T. \label{Gwith}
\end{align}
Equations (\ref{Swith}) and (\ref{Gwith}) reflect two fundamental effects of the GB term on black hole thermodynamics.

In extra-dimensional physics without compactification, the GB coupling constant $\alpha$ is proportional to the inverse string tension with positive coefficient \cite{Boulware:1985wk}, so it is always positive. However, in four dimensions, as the GB term is a topological invariant and does not affect space-time, $\alpha$ is free to be chosen both positive and negative. Actually, it was pointed out that, only if $\alpha$ is allowed to be negative, can there be a reentrant phase transition \cite{sanren}. Furthermore, the positivity of entropy in Eq. (\ref{Swith}) sets a lower bound of $\alpha$, $\alpha>-{(r_+^2+a^2)}/{(4\Xi)}$. Whereas, for positive $\alpha$, if it is below an upper bound, the HP phase transition can happen; if not, no HP phase transition, all to be carefully explained in Sect. \ref{sec:HPGB}.

Last, since $\alpha$ possesses a dimension of $[{\rm length}]^2$, it can be regarded as a thermodynamic variable. From Eqs. (\ref{M}), (\ref{T}), and (\ref{Swith}), the conjugate potential of $\alpha$ is
\begin{align}
X=\lt(\f{\p M}{\p\alpha}\rt)_{S,p,J,Q}=\f{({\p M}/{\p S})_{r_+,p,J,Q}}{({\p\alpha}/{\p S})_{r_+,p,J,Q}}=-4\pi T. \label{Malpha}
\end{align}
Hence, the first law of black hole thermodynamics and the Smarr relation are extended to $\rd M=T\,\rd S+V\,\rd p+\Phi\,\rd Q+\Omega\,\rd J+X\,\rd\alpha$ and $M=2TS-2pV+\Phi Q+2\Omega J+2X\alpha$. Attention, this does not mean that $M$ depends on $\alpha$. The GB term has no dynamic effect in four dimensions, so the black hole mass remains invariant. In fact, the terms $X\,\rd\alpha$ in $\rd M$ and $2X\alpha$ in $M$ exactly compensate the corrections in $S$ in Eq. (\ref{Swith}). Therefore, we can directly obtain the relevant formulae in the GB gravity with $\alpha$, simply by replacing $S$ to $S-4\pi\alpha$ in Eqs. (\ref{M}), (\ref{T}), (\ref{Phi}), and (\ref{Omega}),

\begin{widetext}
\begin{align}
M&=\sqrt{\f{S-4\pi\alpha}{4\pi}\lt\{\lt[1+\f{\pi Q^2}{S-4\pi\alpha}+\f{8p}{3}(S-4\pi\alpha)\rt]^2
+\f{4\pi^2J^2}{(S-4\pi\alpha)^2}\lt[1+\f{8p}{3}(S-4\pi\alpha)\rt]\rt\}}, \label{Mwith}\\
T&=\f{1}{8\pi M}\lt\{\lt[1+\f{\pi Q^2}{S-4\pi\alpha}+\f{8p}{3}(S-4\pi\alpha)\rt]
\lt[1-\f{\pi Q^2}{S-4\pi\alpha}+8p(S-4\pi\alpha)\rt]-\f{4\pi^2J^2}{(S-4\pi\alpha)^2}\rt\},\label{Twith}\\
\Phi&=\f{Q}{2M}\lt[1+\f{\pi Q^2}{S-4\pi\alpha}+\f{8p}{3}(S-4\pi\alpha)\rt],\label{Phiwith}\\
\Omega&=\f{\pi J}{M(S-4\pi\alpha)}\lt[1+\f{8p}{3}(S-4\pi\alpha)\rt]. \label{Omegawith}
\end{align}
\end{widetext}

\section{HP phase transitions in the GB gravity} \label{sec:HPGB}

In this section, we study the HP phase transitions of the Schwarzschild--AdS, RN--AdS, and KN--AdS black holes in the GB gravity in order. In each case, we discuss the relevant issues first without and then with the GB term. For the simple Schwarzschild--AdS and RN--AdS black holes, analytical solutions are given, so as to present clear mathematical description. However, for the complicated KN--AdS black, only numerical results are shown diagrammatically for intuitive physical comprehension.

The basic strategy of our calculations consists of two steps. First, at the HP phase transition point, we substitute Eqs. (\ref{Mwith})--(\ref{Omegawith}) into the criterion in Eq. (\ref{panju}) to obtain the black hole entropy $S$ in terms of $p$, $\Phi$, $\Omega$, and $\alpha$. By this means, the HP phase transition temperature $T_{\rm HP}$ can be determined. Second, we solve $S$ from Eq. (\ref{Twith}) as a function of $T$ and substitute it into Eq. (\ref{G}) to obtain the black hole Gibbs free energy at arbitrary temperature and pressure. Compared with the vanishing Gibbs free energy of the thermal AdS space, the global phase structures of the HP phase transitions can finally be achieved. In brief, we utilize the black hole entropy $S$ as the intermediate variable in all calculations.

We should stress that this method is different from the frequently-used ones in the literature (i.e., express everything as a function of the horizon radius $r_+$). The difference between these two methods is not evident for the Schwarzschild--AdS and RN--AdS black holes with simple spherical horizons, but is essential for the KN--AdS black holes without spherical horizons. In this circumstance, the calculations via $r_+$ are usually rather tedious and even problematic, and the method via $S$ will prove systematic and efficient.

\subsection{HP phase transitions of the Schwarzschild--AdS black holes} \label{sec:1}

The HP phase transitions of the Schwarzschild--AdS black holes in the GB gravity have been explored in the literature. However, there still remains some subtlety to be clarified, so we discuss them both for completeness and as a demonstration of our calculational method. The same procedure will be applied to the more complicated black hole solutions, and the results below will be taken for comparison in Sects. \ref{sec:2} and \ref{sec:3}.

\begin{figure}[h]
\begin{center}
\includegraphics[width=0.95\linewidth,angle=0]{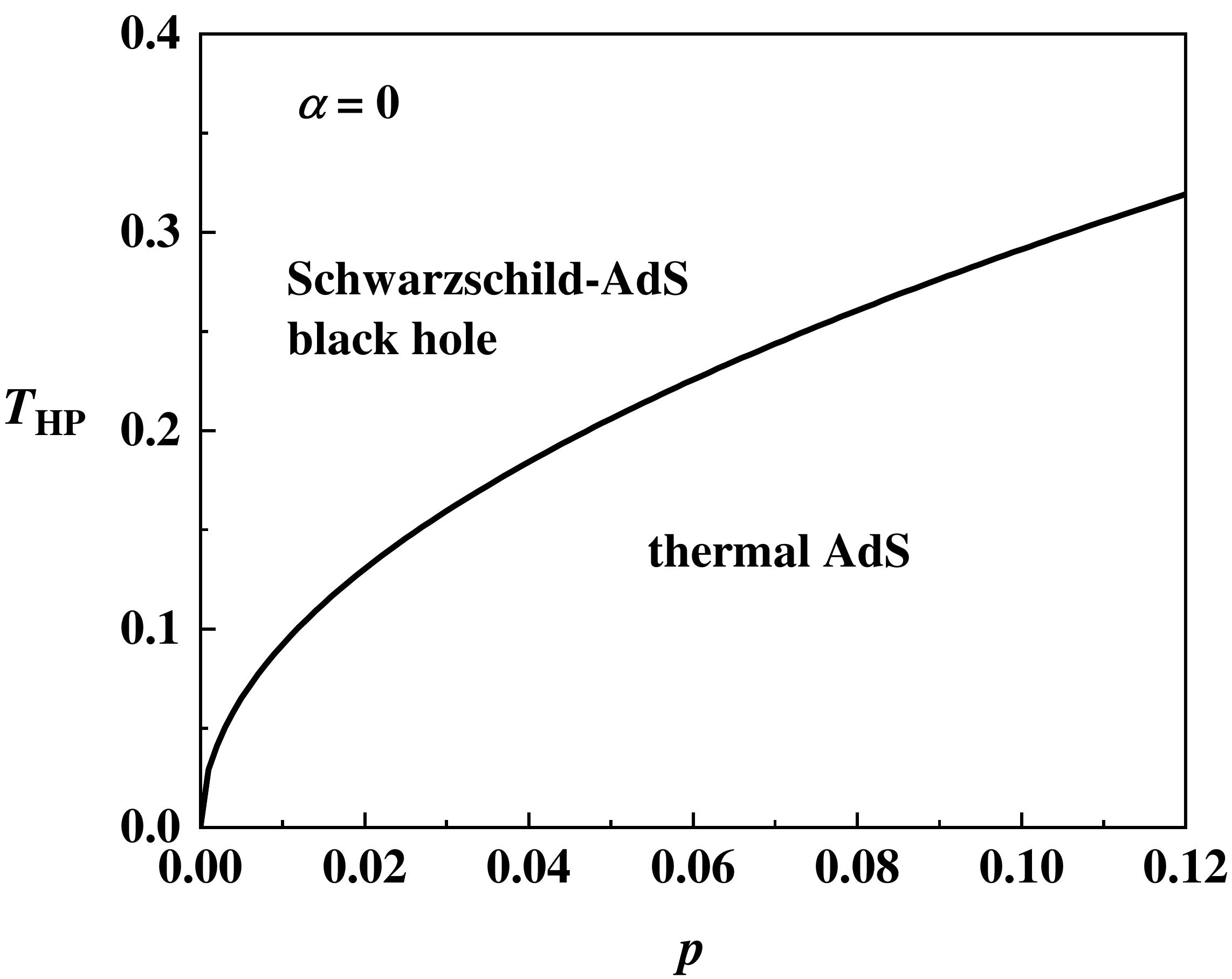}
\end{center}
\caption{The HP phase transition temperature of the Schwarzschild--AdS black holes as a function of pressure. There is no terminal point in the coexistence line, and the HP phase transition can happen at all pressures. The thermal AdS phase lies below the coexistence line, acting as a solid in the HP phase transition.} \label{f:TPSAdSwithout}
\end{figure}

First, we begin our discussions without the GB term. For the Schwarzschild--AdS black holes, Eqs. (\ref{Mwith}) and (\ref{Twith}) are reduced to
\begin{align}
M&=\sqrt{\f{S}{4\pi}}\lt(1+\f{8pS}{3}\rt), \label{MSAdSwithout}\\
T&=\f{1}{\sqrt{16\pi S}}(1+8pS), \label{TSAdSwithout}
\end{align}
so the Gibbs free energy is
\begin{align}
G=M-TS=\sqrt{\f{S}{16\pi}}\lt(1-\f{8pS}{3}\rt).\label{GS}
\end{align}
Hence, at the HP phase transition point, from the criterion in Eq. (\ref{panju}), we obtain $S={3}/{(8p)}$. Substituting it into Eq. (\ref{TSAdSwithout}), the HP phase transition temperature is
\begin{align}
T_{\rm HP}=\sqrt{\f{8p}{3\pi}}. \label{TpSAdSwithout}
\end{align}
Naturally, this is just the equation of coexistence line in the $T$--$p$ phase diagram, as shown in Fig. \ref{f:TPSAdSwithout}. Since $p$ has no bound in Eq. (\ref{TpSAdSwithout}), there is no terminal point in the coexistence line, and the HP phase transition can happen at all pressures, without a critical point. Therefore, it is more like a solid--liquid phase transition, rather than a liquid--gas phase transition. Interestingly, the thermal AdS phase even plays the role of a solid, as it lies below the coexistence line \cite{rev3}.

\begin{figure}[h]
\begin{center}
\includegraphics[width=0.95\linewidth,angle=0]{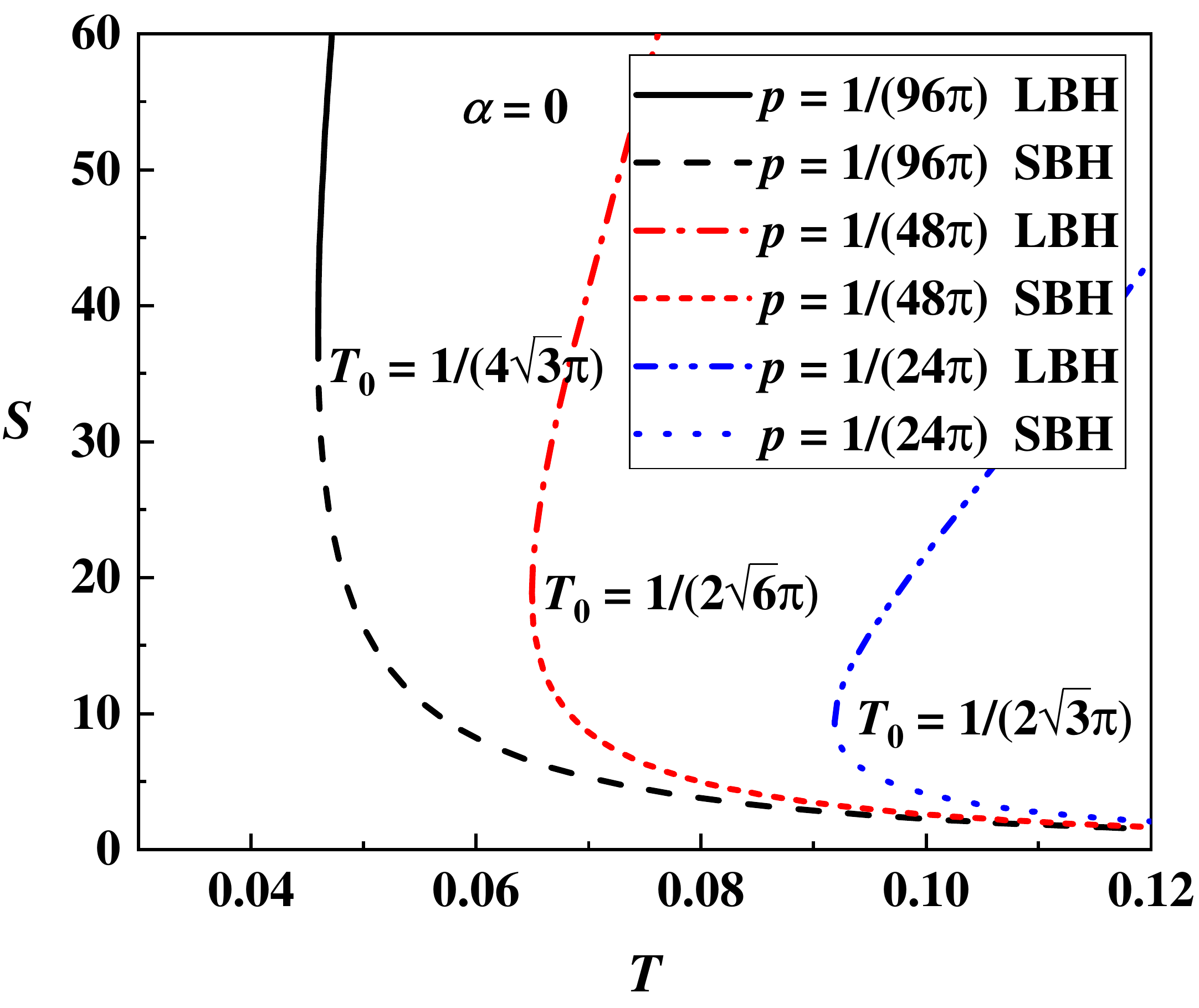}
\end{center}
\caption{The entropies of the Schwarzschild--AdS black holes as a function of temperature (LBH and SBH stand for large and small black holes). The large black holes have positive heat capacities and are thermodynamically stable, and small ones are on the contrary. At a given pressure $p$, the black hole temperature has a positive minimum, $T_0=\sqrt{2p/\pi}$.} \label{f:STSAdSwithout}
\end{figure}

Next, from Eq. (\ref{TSAdSwithout}), we can solve $S$ in terms of $T$ and $p$,
\begin{align}
S(T,p)=\f{1}{8p^2}\lt(\pi T^2-p \pm T\sqrt{\pi^2 T^2-2\pi p}\rt),\label{STSAdSwithout}
\end{align}
where $\pm$ correspond to large and small black holes respectively (with large and small entropies and horizon radii). The $S$--$T$ curves are plotted in Fig. \ref{f:STSAdSwithout}. As the heat capacity at constant pressure is $C_p=T(\p S/\p T)_p$, from the slopes of the $S$--$T$ curves, we see that large black holes are thermodynamically stable with positive $C_p$, but small ones are unstable with negative $C_p$ and thus cannot establish the equilibrium with the thermal AdS space. Moreover, from Eq. (\ref{STSAdSwithout}), the Schwarzschild--AdS black hole temperature must have a positive minimum,
\begin{align}
T_0=\sqrt{\f{2p}{\pi}}. \n
\end{align}
Actually, $(T_0,S(T_0))=(\sqrt{2p/\pi},1/(8p))$ is just the meeting point of the $S$--$T$ curves for large and small black holes in Fig. \ref{f:STSAdSwithout}.

\begin{figure}[h]
\begin{center}
\includegraphics[width=0.95\linewidth,angle=0]{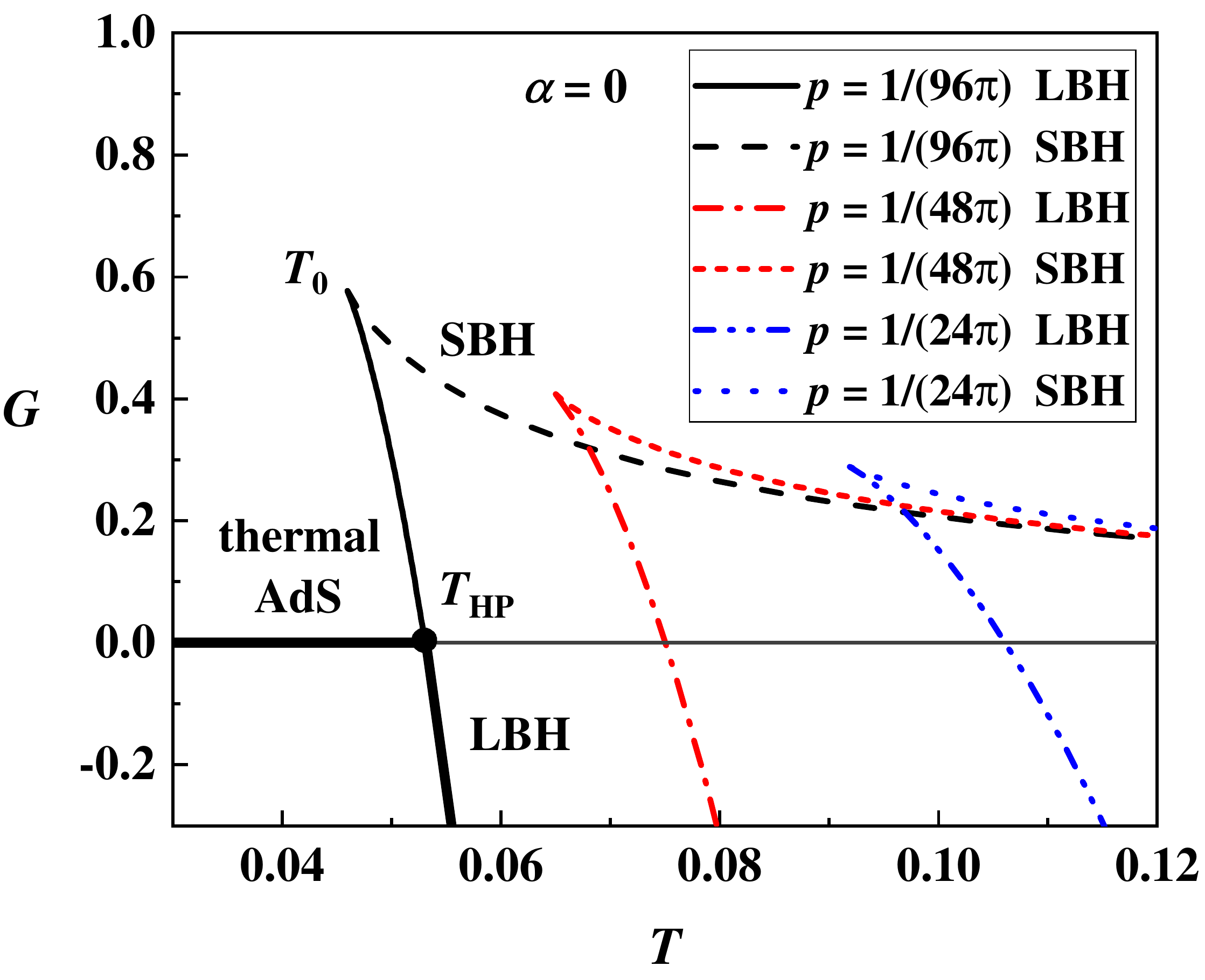}
\end{center}
\caption{The Gibbs free energies of large and small Schwarzschild--AdS black holes at arbitrary temperature and pressure. The $G$--$T$ curves of the thermal AdS space and the stable large black holes intersect at the HP phase transition temperature $T_{\rm HP}$, which increases at large pressures. Below or above $T_{\rm HP}$, the thermal AdS or the large black hole phase is globally preferred. The $G$--$T$ curves of the unstable small black holes are always above the $T$-axis, so the HP phase transition never happens.} \label{f:GTSAdSwithout}
\end{figure}

Substituting Eq. (\ref{STSAdSwithout}) into (\ref{GS}), the Gibbs free energies of large and small Schwarzschild--AdS black holes at arbitrary temperature and pressure are obtained as
\begin{align}
G(T,p)&=\frac{\sqrt{\pi T^2-p\pm T\sqrt{\pi^2 T^2-2\pi p}}}{24\sqrt{2\pi}p^2} \n\\
&\quad\times\lt(4p- \pi T^2\mp T\sqrt{\pi^2 T^2-2\pi p}\rt).\label{GTSAdSwithout}
\end{align}
The $G$--$T$ curves are shown in Fig. \ref{f:GTSAdSwithout}. The two branches of the curves correspond to large and small black holes, meeting with a cusp at $(T_0,G(T_0))=(\sqrt{2p/\pi},1/(12\sqrt{2\pi p}))$. We observe that both Gibbs free energies of large and small black holes decrease with temperature. For the unstable small black holes, the $G$--$T$ curves are concave and will never reach the $T$-axis ($G$ only tends to 0 when $T\to\infty$), so there is no HP phase transition. On the contrary, for the stable large black holes, their $G$--$T$ curves are convex and will cross the $T$-axis at the HP phase transition temperatures $T_{\rm HP}$. With $p$ increasing, $T_{\rm HP}$ moves rightward, indicating that $T_{\rm HP}$ increases at high pressures, consistent with Fig. \ref{f:TPSAdSwithout}. Setting $G=0$ in Eq. (\ref{GTSAdSwithout}), we recover the result of $T_{\rm HP}$ in Eq. (\ref{TpSAdSwithout}). Below $T_{\rm HP}$, the vanishing Gibbs free energy of the thermal AdS space is lower than that of the large black hole; above $T_{\rm HP}$, the Gibbs free energy of the large black hole becomes negative and is thus lower than that of the thermal AdS space. As a result, there is a discontinuity in the first order derivatives of the Gibbs free energies of the black hole--thermal AdS system, corresponding to a first order phase transition at $T_{\rm HP}$.

Since the unstable small black holes have no HP phase transition and more importantly cannot be in equilibrium with the thermal AdS space at all, we will omit the related discussions on them in the GB gravity below.

Now, we take into account the GB term and discuss its effects on the HP phase transitions. In the GB gravity, from Eqs. (\ref{Mwith}) and (\ref{Twith}), the mass and temperature of the Schwarzschild--AdS black hole should be modified to
\begin{align}
M&=\sqrt{\f{S-4\pi\alpha}{4\pi}}\lt[1+\f{8p}{3}(S-4\pi\alpha)\rt], \label{MSAdSwith}\\
T&=\f{1}{\sqrt{16\pi(S-4\pi\alpha)}} \lt[1+8p(S-4\pi\alpha)\rt], \label{TSAdSwith}
\end{align}
so the Gibbs free energy becomes
\begin{align}
G&=\sqrt{\frac{S-4\pi\alpha}{16\pi}}\lt(\frac{S-8\pi\alpha}{S-4\pi\alpha}-\frac{8pS}{3}-\f{64\pi\alpha p}{3}\rt).\label{GSAdSwith}
\end{align}
Therefore, at the HP phase transition point, for large black holes, from Eq. (\ref{GSAdSwith}), we have
\begin{align}
S=\f{1}{16p}\lt(3-32\pi\alpha p+\sqrt{9-960\pi\alpha p+9216\pi^2\alpha^2 p^2}\rt). \n
\end{align}
Substituting it into Eq. (\ref{TSAdSwith}), we obtain the HP phase transition temperature as
\begin{align}
T_{\rm HP}=\sqrt{\f{p}{4\pi}}\f{5-96\pi\alpha p+\sqrt{9-960\pi\alpha p+9216\pi^2\alpha^2 p^2}}
{\sqrt{3-96\pi\alpha p+\sqrt{9-960\pi\alpha p+9216\pi^2\alpha^2 p^2}}}. \label{TpSAdSwith}
\end{align}
This result is naturally reduced to Eq. (\ref{TpSAdSwithout}) if $\alpha$ vanishes. Moreover, $T_{\rm HP}$ is a decreasing function of $\alpha$. Thus, if the HP phase transition can happen, $\alpha$ should have an upper bound, such that $T_{\rm HP}>T_0$, to be explained in more detail below.

Furthermore, from Eqs. (\ref{TSAdSwith}) and (\ref{STSAdSwithout}), we can solve $S$ in terms of $T$, $p$, and $\alpha$,
\begin{align}
S(T,p,\alpha)=S(T,p)+4\pi\alpha, \label{SSAdSwitha}
\end{align}
so there is only a shift $4\pi\alpha$ in entropy, as expected in Eq. (\ref{Swith}). Besides, the minimal black hole temperature, $T_0=\sqrt{2p/\pi}$, remains unchanged in the presence of $\alpha$.

Then, substituting Eq. (\ref{SSAdSwitha}) into (\ref{GSAdSwith}) and using Eq. (\ref{GTSAdSwithout}), the Gibbs free energy of large Schwarzschild--AdS black holes with the GB term is
\begin{align}
G(T,p,\alpha)=G(T,p)-4\pi\alpha T. \label{www}
\end{align}
Therefore, $G$ decreases with $\alpha$, leading to a lower $T_{\rm HP}$.

Before plotting the coexistence lines and the $G$--$T$ curves for the Schwarzschild--AdS black holes with the GB term, a crucial issue must be carefully elucidated, that is, the lower and upper bounds of the GB coupling constant $\alpha$. The bounds of $\alpha$ come from the two corrections by the GB term to the black hole entropy and Gibbs free energy: $S+4\pi\alpha$ and $G-4\pi\alpha T$. On the one hand, $\alpha$ cannot be too negative, otherwise the minimum of $S$ would be negative, in contradiction to the positivity of entropy; on the other hand, if the HP phase transition happens, $\alpha$ cannot be too positive either, otherwise the maximum of $G$ would be negative, even lower than the Gibbs free energy of the thermal AdS space. \footnote{Of course, we do not state that there is some physical principle that would guarantee the existence of the HP phase transition. We only mean that, if the HP phase transition happens, $\alpha$ must have lower and upper bounds at the same time.} First, from Eq. (\ref{SSAdSwitha}), $S> S(T_0)=1/(8p)+4\pi\alpha> 0$, so $\alpha$ has a lower bound, $\alpha> -1/(32\pi p)$, consistent with the condition $\alpha> -r_+^2/4$ in Sect. \ref{sec:GB}. Second, because the Gibbs free energy is a monotonically decreasing function of temperature, the maximum of $G$ should be evaluated as $G(T_0)=1/(12\sqrt{2\pi p})-4\sqrt{2\pi p}\alpha$. This value should be positive if there exists the HP phase transition, so it sets an upper bound of $\alpha$, $\alpha< 1/(96\pi p)$. Altogether, in the presence of the HP phase transition, $\alpha$ has both lower and upper bounds simultaneously,
\begin{align}
-\f{1}{32\pi p}<\alpha< \f{1}{96\pi p}. \label{alphaxian}
\end{align}
Moreover, the bounds in Eq. (\ref{alphaxian}) automatically satisfy the requirements that the terms inside the square roots must be positive in Eq. (\ref{TpSAdSwith}).

\begin{figure}[h]
\begin{center}
\includegraphics[width=0.95\linewidth,angle=0]{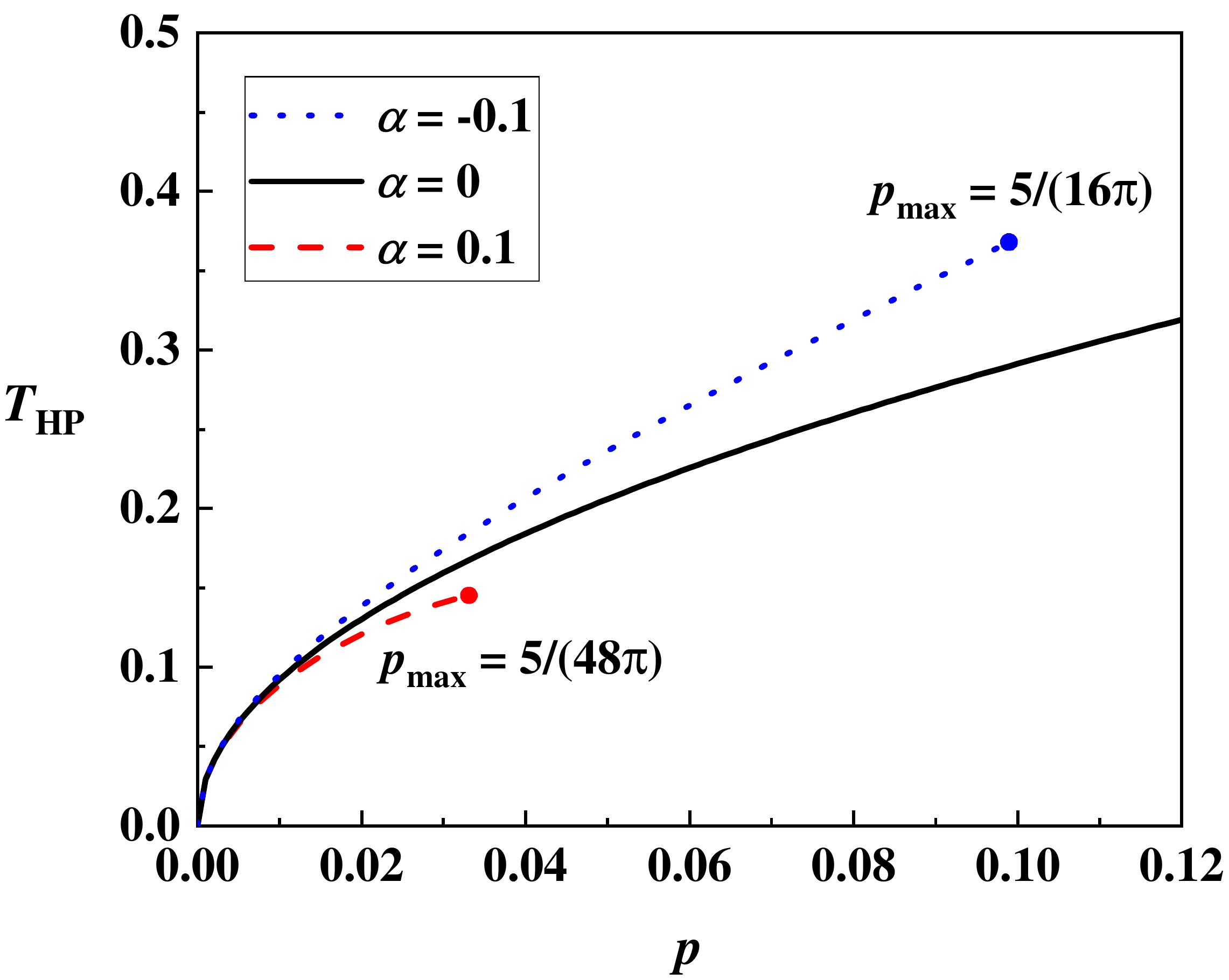}
\end{center}
\caption{The HP phase transition temperature of the Schwarzschild--AdS black holes with the GB term as a function of pressure. At a fixed pressure $p$, $T_{\rm HP}$ decreases with $\alpha$. There is an upper bound of $p$, $p< 1/(96\pi\alpha)$ for positive $\alpha$ or $p<-1/(32\pi\alpha)$ for negative $\alpha$, resulting in the terminal points in the coexistence lines.} \label{f:TpSAdSwith}
\end{figure}

The bounds of $\alpha$ cause significant difference between the coexistence lines of the Schwarzschild--AdS black holes without and with the GB term. From Eq. (\ref{alphaxian}),
\begin{align}
p&<p_{\rm max}=\f{1}{96\pi\alpha} \qquad ({\rm for}~\alpha>0), \n\\
p&<p_{\rm max}=-\f{1}{32\pi\alpha} \qquad ({\rm for}~\alpha<0). \label{pxian}
\end{align}
Therefore, unless $\alpha=0$, no matter how positive or negative it is, there is a corresponding upper bound of pressure. As a result, there must be terminal points in the coexistence lines, and the HP phase transitions can happen only below the critical pressures $p_{\rm max}$. From Eq. (\ref{TpSAdSwith}), the coexistence lines of the Schwarzschild--AdS black holes with the GB term are plotted in Fig. \ref{f:TpSAdSwith}, with different values of $\alpha$. Moreover, we find that $T_{\rm HP}$ decreases with $\alpha$ at a fixed pressure.

\begin{figure}[h]
\begin{center}
\includegraphics[width=0.95\linewidth,angle=0]{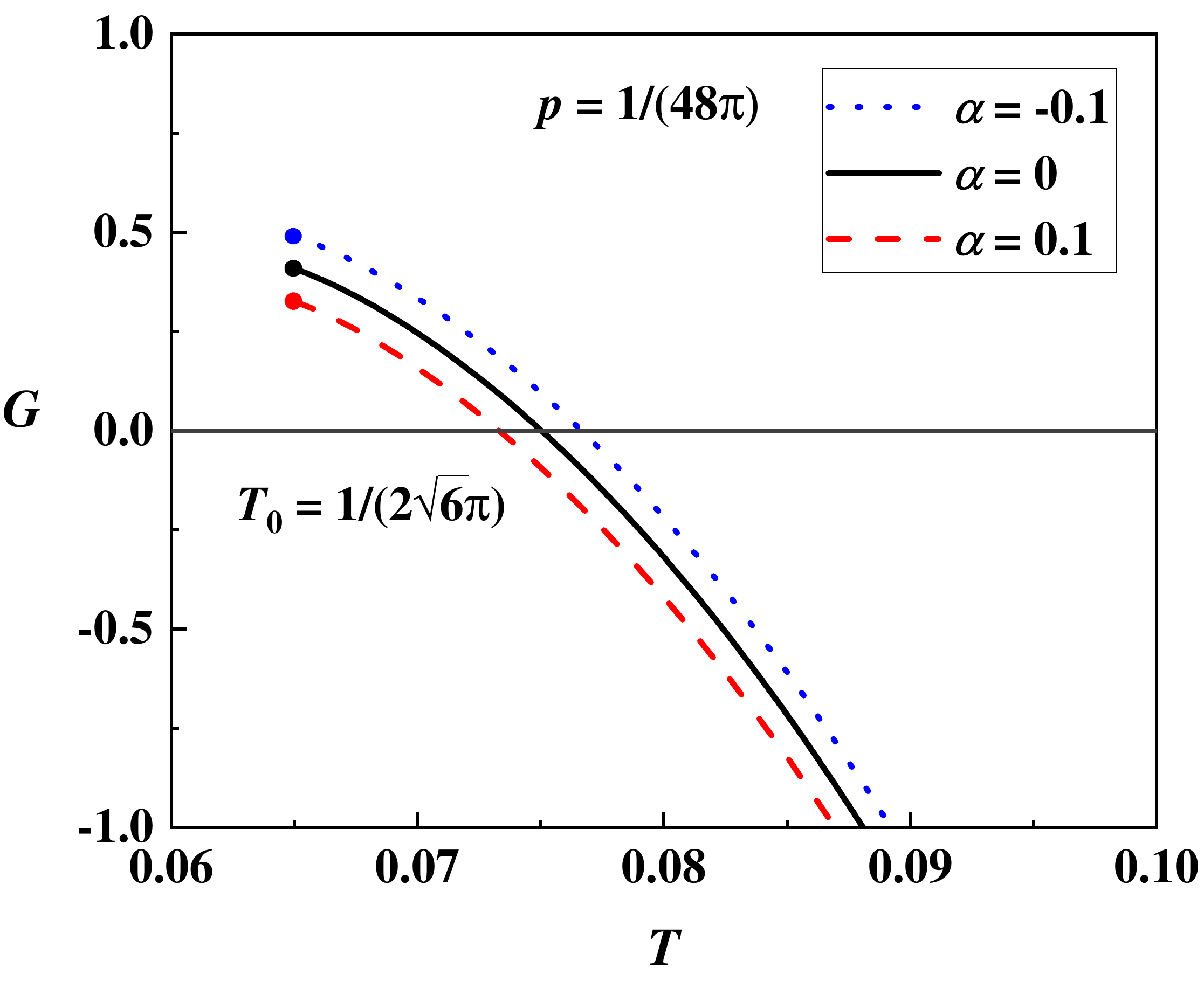}
\end{center}
\caption{The Gibbs free energy of large Schwarzschild--AdS black holes as a function of temperature, with a fixed pressure $p=1/(48\pi)$ and different values of $\alpha$. All the $G$--$T$ curves start from the same minimal temperature $T_0=1/(2\sqrt{6}\pi)$ and move downward with $\alpha$ increasing, so $T_{\rm HP}$ decreases with $\alpha$.} \label{f:GTSAdSwith}
\end{figure}

Last, the $G$--$T$ curves of large Schwarzschild--AdS black holes with the GB term are shown in Fig. \ref{f:GTSAdSwith}, with different values of $\alpha$. According to Eq. (\ref{www}), all these curves set off from the same minimal temperature $T_0=\sqrt{2p/\pi}$, and the effect of the GB term is just to proportionally translate the $G$--$T$ curves downward when $\alpha$ increases, inducing a lower $T_{\rm HP}$, consistent with Fig. \ref{f:TpSAdSwith}.

\subsection{HP phase transitions of the RN--AdS black holes} \label{sec:2}

We continue to study the HP phase transitions of the RN--AdS black holes in the GB gravity. This issue was mentioned in Ref. \cite{sanren}, but the corresponding discussions were actually absent, so we reconsider this problem in more detail. In principle, the procedure is in parallel to that in Sect. \ref{sec:1}, but this does not mean that the whole process is merely a repetition. There is an intrinsic difference in between. Due to the conservation of charge, the RN--AdS black hole with fixed charge cannot undergo the HP phase transition to the thermal AdS space without charge. Consequently, the discussion of the HP phase transition of the RN--AdS black holes should performed in the grand canonical ensemble, in which the electric potential $\Phi$ is fixed and the charge $Q$ is allowed to change.

From Eq. (\ref{dugui}), the metric of the RN--AdS black hole is
\begin{align}
\rd s^2=-f(r)\,\rd t^2+\f{\rd r^2}{f(r)}+r^2\,\rd\theta^2+r^2\sin^2\theta\,\rd\phi^2,\n
\end{align}
with $f(r)=1-2M/r+Q^2/r^2+8\pi pr^2/3$. The horizon radius is determined by $f(r_+)=0$, and in the limit of a vanishing cosmological constant (i.e., $p\to 0$), $r_+=M+\sqrt{M^2-Q^2}$. Hence, we have $Q<M$, so the electric potential at the horizon must satisfy
\begin{align}
\Phi=\f{Q}{r_+}=\f{Q}{M+\sqrt{M^2-Q^2}}<\f{Q}{M}<1.\n
\end{align}
Therefore, we focus on the fixed electric potential ensemble with $\Phi<1$ in the following discussions.

\begin{figure}[h]
\begin{center}
\includegraphics[width=0.95\linewidth,angle=0]{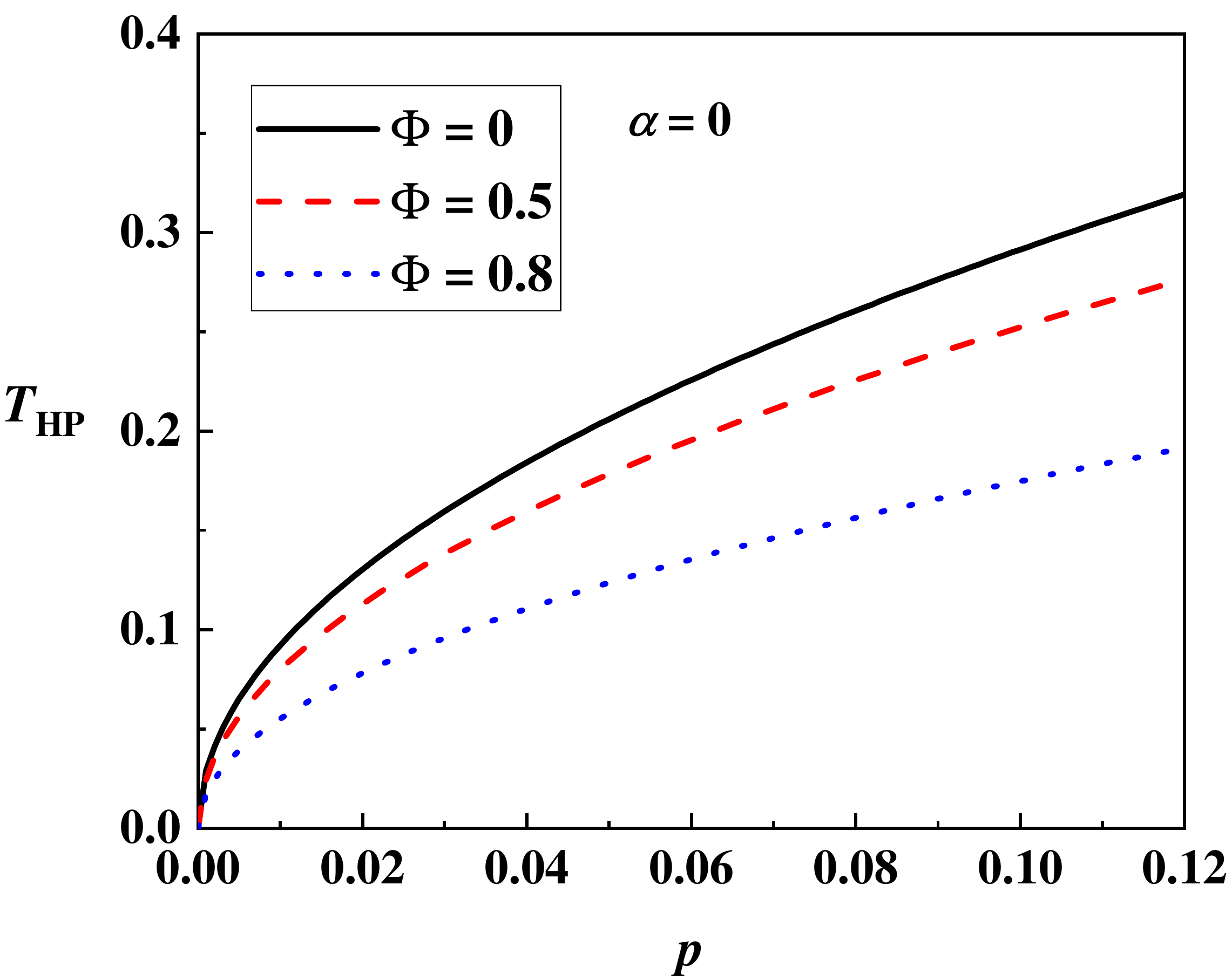}
\end{center}
\caption{The HP phase transition temperature of the RN--AdS black holes as a function of pressure, with different values of $\Phi$. The HP phase transition can happen at all pressures, and $T_{\rm HP}$ decreases with $\Phi$ at a fixed pressure.} \label{f:TpRNAdSwithout}
\end{figure}

First, for the RN--AdS black holes without the GB term, from Eqs. (\ref{Mwith})--(\ref{Phiwith}), we have
\begin{align}
M&=\sqrt{\frac{S}{4\pi}}\lt(1+\Phi^2+\f{8pS}{3}\rt),\label{MRNAdSwithout}\\
T&=\f{1}{\sqrt{16\pi S}}(1-\Phi^2+8pS),\label{TRNAdSwithout}\\
\Phi&=Q\sqrt{\f{\pi}{S}},\label{PhiRNAdSwithout}
\end{align}
where the variable $Q$ in $M$ and $T$ has been replaced to $\Phi$ by virtue of Eq. (\ref{PhiRNAdSwithout}). Therefore, the Gibbs free energy of the RN--AdS black holes in the grand canonical ensemble reads
\begin{align}
G=M-TS-\Phi Q=\sqrt{\f{S}{16\pi}}\lt(1-\Phi^2-\f{8pS}{3}\rt).\label{GRN}
\end{align}
At the HP phase transition point, from Eq. (\ref{GRN}), we have $S=3(1-\Phi^2)/(8p)$. Substituting it into Eq. (\ref{TRNAdSwithout}), we obtain the HP phase transition temperature of the RN--AdS black holes as
\begin{align}
T_{\rm HP}=\sqrt{\f{8p}{3\pi}(1-\Phi^2)}. \label{TpQRNAdSwithout}
\end{align}
This result naturally is reduced to Eq. (\ref{TpSAdSwithout}) if $\Phi$ or $Q$ vanishes, and again there is no bound of $p$, so the HP phase transition can happen at all pressures. The coexistence lines are shown in Fig. \ref{f:TpRNAdSwithout}, with different values of $\Phi$, and at a given pressure, $T_{\rm HP}$ is found to decrease with $\Phi$.

\begin{figure}[h]
\begin{center}
\includegraphics[width=0.95\linewidth,angle=0]{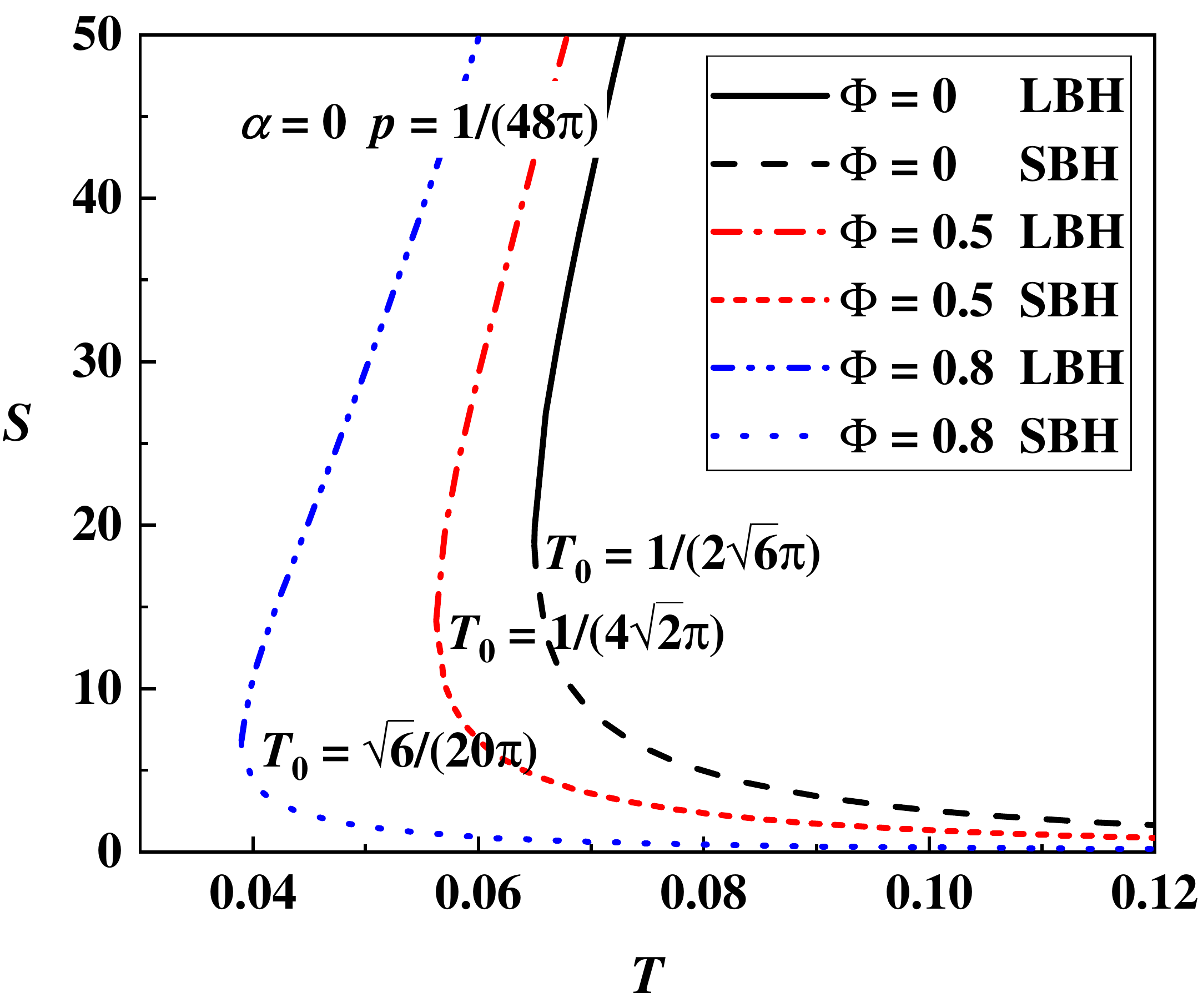}
\end{center}
\caption{The entropies of large and small RN--AdS black holes as a function of temperature, with a fixed pressure $p=1/(48\pi)$ and different values of $\Phi$. Similar to the Schwarzschild--AdS black holes with the minimal temperatures $T_0=\sqrt{2p/\pi}$, the RN--AdS black hole temperature also has its minimum $T_0=\sqrt{{2p}(1-\Phi^2)/\pi}$, corrected by the electric potential.} \label{f:STRNAdSwithout}
\end{figure}

Furthermore, from Eq. (\ref{TRNAdSwithout}), we can again solve $S$ in terms of $T$, $p$, and $\Phi$,
\begin{align}
S(T,p,\Phi)&=\f{1}{8p^2}\Big[\pi T^2-p(1-\Phi^2) \n\\
&\quad \pm T\sqrt{\pi^2 T^2-2\pi p(1-\Phi^2)}\Big].\label{SRNAdSwithout}
\end{align}
Then, at a given pressure $p$, there is also a minimum of the RN--AdS black hole temperature,
\begin{align}
T_0=\sqrt{\f{2p}{\pi}(1-\Phi^2)},\label{t0}
\end{align}
and the $S$--$T$ curves of the RN--AdS black holes are shown in Fig. \ref{f:STRNAdSwithout}, which are qualitatively similar to those of the Schwarzschild--AdS black holes in Fig. \ref{f:STSAdSwithout}.

\begin{figure}[h]
\begin{center}
\includegraphics[width=0.95\linewidth,angle=0]{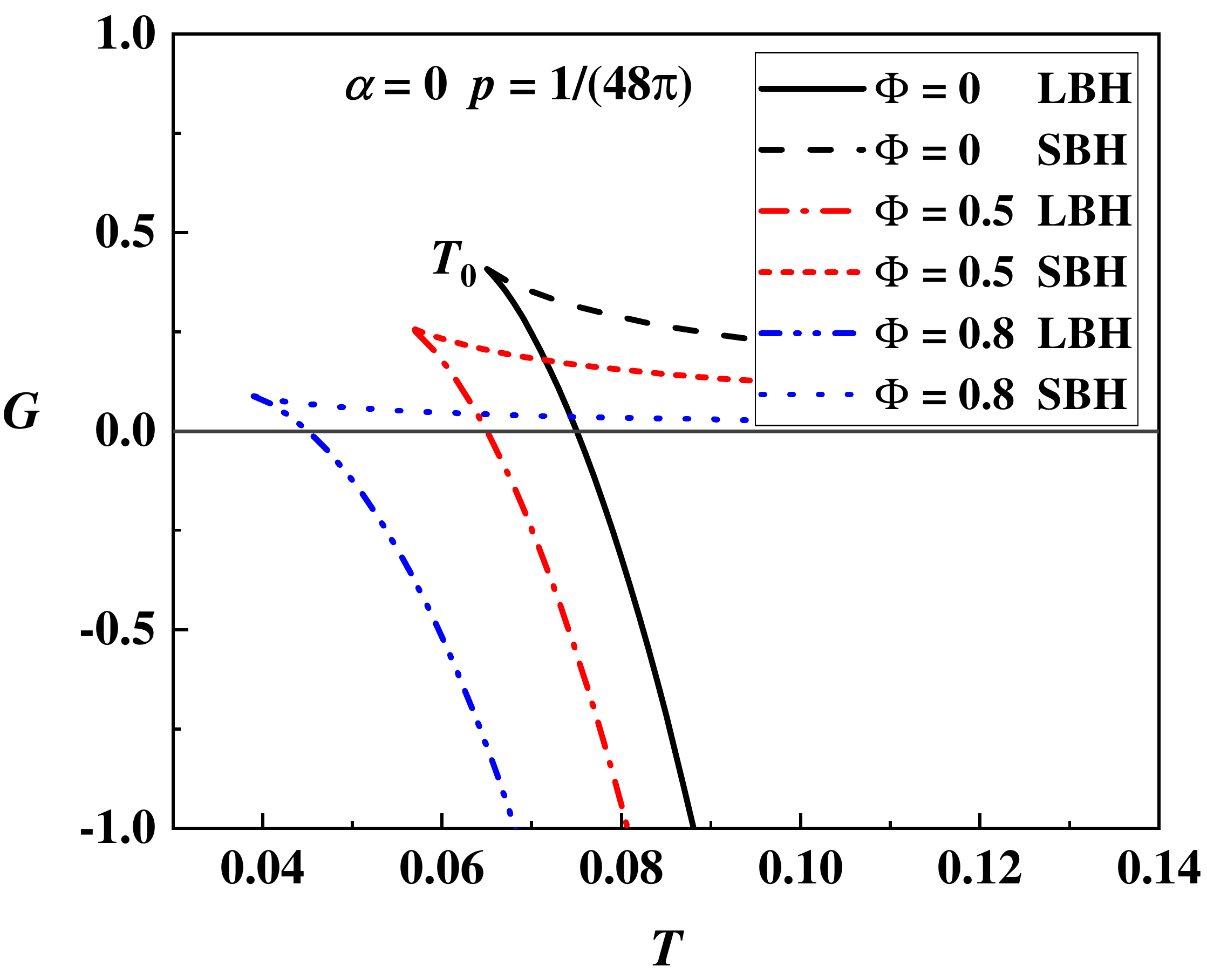}
\end{center}
\caption{The Gibbs free energy of the RN--AdS black holes as a function of temperature, with a fixed pressure $p=1/(48\pi)$ and different values of $\Phi$. $T_{\rm HP}$ decreases with $\Phi$, and the RN--AdS black holes have minimal temperatures $T_0$.} \label{f:GTRNAdSwithout}
\end{figure}

Next, substituting Eq. (\ref{SRNAdSwithout}) into (\ref{GRN}), the Gibbs free energy of the RN--AdS black holes at arbitrary temperature and pressure in the grand canonical ensemble can be obtained straightforwardly,

\begin{widetext}
\begin{align}
G(T,p,\Phi)=\frac{\sqrt{\pi T^2-p(1-\Phi^2)\pm T\sqrt{\pi^2 T^2-2\pi p(1-\Phi^2)}}}{24\sqrt{2\pi}p^2}\lt[4p(1-\Phi^2)- \pi T^2\mp T\sqrt{\pi^2 T^2-2\pi p(1-\Phi^2)}\rt],\n
\end{align}
\end{widetext}
which is a natural extension of Eq. (\ref{GTSAdSwithout}). The $G$--$T$ curves are plotted in Fig. \ref{f:GTRNAdSwithout}, with different values of $\Phi$. We clearly observe two features: first, the HP phase transition temperature $T_{\rm HP}$ decreases with $\Phi$, as expected in Eq. (\ref{TpQRNAdSwithout}); second, the RN--AdS black hole temperature $T$ also has a minimum $T_0$, as explained above.

Below, for the HP phase transition of the RN--AdS black holes with the GB term, we repeat the above calculations, by replacing $S$ to $S-4\pi\alpha$ in Eqs. (\ref{MRNAdSwithout})--(\ref{PhiRNAdSwithout}), and the corresponding results of $M$, $T$, and $\Phi$ are
\begin{align}
M&=\sqrt{\frac{S-4\pi\alpha}{4\pi}}\lt[1+\Phi^2+\f{8p}{3}(S-4\pi\alpha)\rt], \n\\
T&=\f{1}{\sqrt{16\pi (S-4\pi\alpha)}}[1-\Phi^2+8p(S-4\pi\alpha)],\n\\
\Phi&=Q\sqrt{\f{\pi}{S-4\pi\alpha}}. \n
\end{align}
Then, the Gibbs free energy reads
\begin{align}
G=\sqrt{\frac{S-4\pi\alpha}{16\pi}}\lt[\frac{S-8\pi\alpha}{S-4\pi\alpha}(1-\Phi^2)-\frac{8pS}{3}-\f{64\pi\alpha p}{3}\rt], \n
\end{align}
Hence, the HP phase transition temperature can be extracted as before,

\vskip 1cm
\begin{widetext}
\begin{align}
T_{\rm HP}&=\sqrt{\frac{p}{4\pi}}\frac{5(1-\Phi^2)-96\pi\alpha p+\sqrt{9(1-\Phi^2)^2-960\pi\alpha p(1-\Phi^2)+9216\pi^2\alpha^2 p^2}}{\sqrt{3(1-\Phi^2)-96\pi\alpha p+\sqrt{9(1-\Phi^2)^2-960\pi\alpha p(1-\Phi^2)+9216\pi^2\alpha^2 p^2}}}. \n
\end{align}
\end{widetext}

Furthermore, expressing $S$ in terms of $T$, $p$, $\Phi$, and $\alpha$,
\begin{align}
S(T,p,\Phi,\alpha)=S(T,p,\Phi) +4\pi\alpha,\n
\end{align}
we finally arrive at the Gibbs free energy of the RN--AdS black holes with the GB term in the grand canonical ensemble,
\begin{align}
G(T,p,\Phi,\alpha)=G(T,p,\Phi)-4\pi\alpha T.\n
\end{align}
The explicit expressions of $S(T,p,\Phi,\alpha)$ and $G(T,p,\Phi,\alpha)$ are omitted due to their unnecessary formal complexities.

Again, before plotting the coexistence lines and the $G$--$T$ curves for the RN--AdS black holes with the GB term, the bounds of $\alpha$ must be determined with care. Similar to the case of the Schwarzschild--AdS black holes, if the HP phase transition happens, there are both lower and upper bounds of $\alpha$, which can easily be obtained via the same analysis in Sect. \ref{sec:1},
\begin{align}
-\f{1-\Phi^2}{32\pi p}<\alpha< \f{1-\Phi^2}{96\pi p}. \n
\end{align}

\begin{figure}[h]
\begin{center}
\includegraphics[width=0.95\linewidth,angle=0]{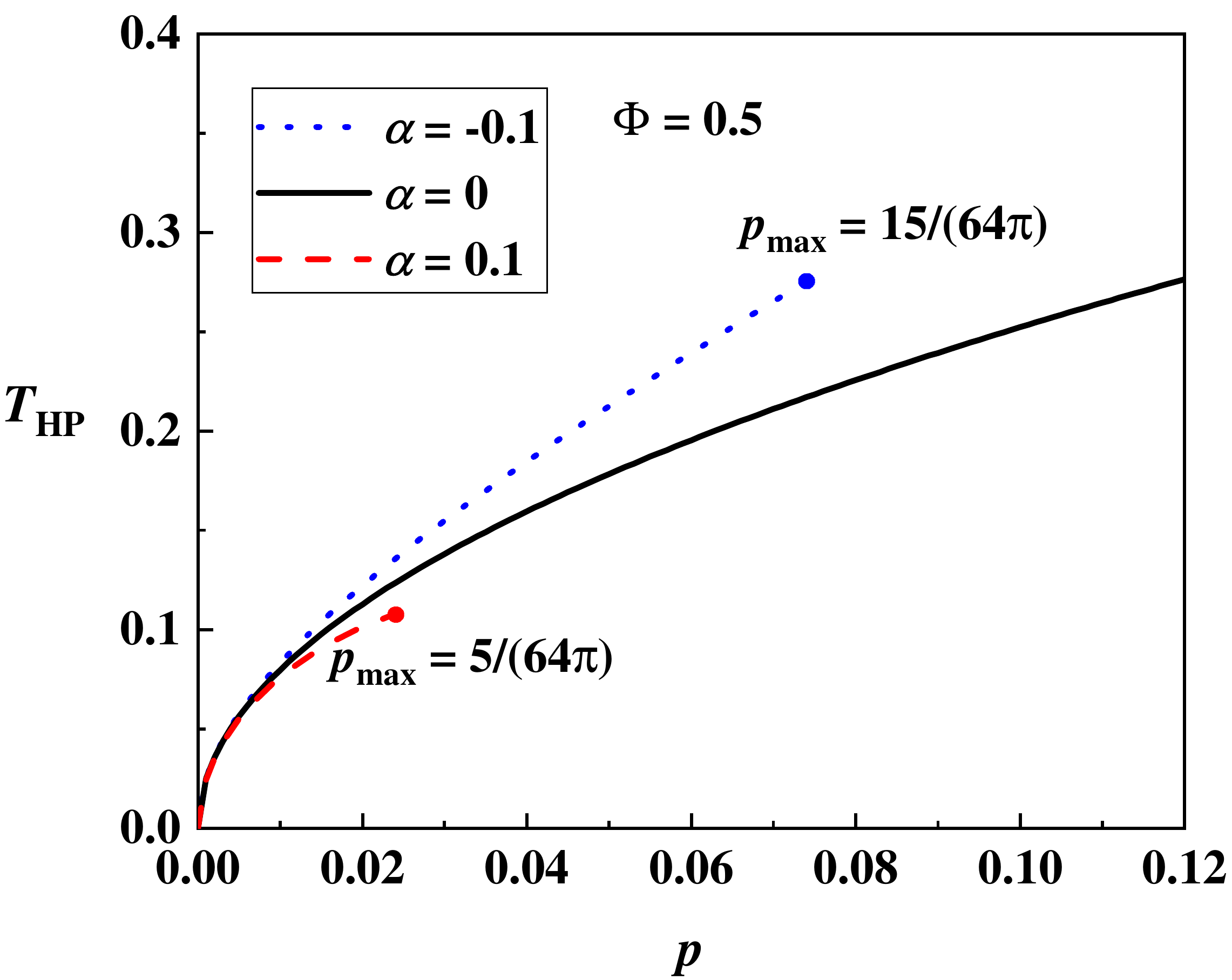}
\end{center}
\caption{The HP phase transition temperature of the RN--AdS black holes with the GB term as a function of pressure, with $\Phi=0.5$ and different values of $\alpha$. $T_{\rm HP}$ decreases with $\alpha$ at a given pressure, and there are terminal points in the coexistence lines, both as same as the Schwarzschild--AdS case in Fig. \ref{f:TpSAdSwith}.} \label{f:TpRNAdSwith}
\end{figure}

The coexistence lines of the RN--AdS black holes are shown in Fig. \ref{f:TpRNAdSwith}, with $\Phi=0.5$ and different values of $\alpha$. We see that $T_{\rm HP}$ decreases with $\alpha$ at a given pressure, and this trend is the same as the Schwarzschild--AdS case. Also, there are terminal points in the coexistence lines, with the critical pressures, $p_{\rm max}=(1-\Phi^2)/(96\pi\alpha)$ for $\alpha>0$ and $p_{\rm max}=-(1-\Phi^2)/(32\pi\alpha)$ for $\alpha<0$.

\begin{figure}[h]
\begin{center}
\includegraphics[width=0.95\linewidth,angle=0]{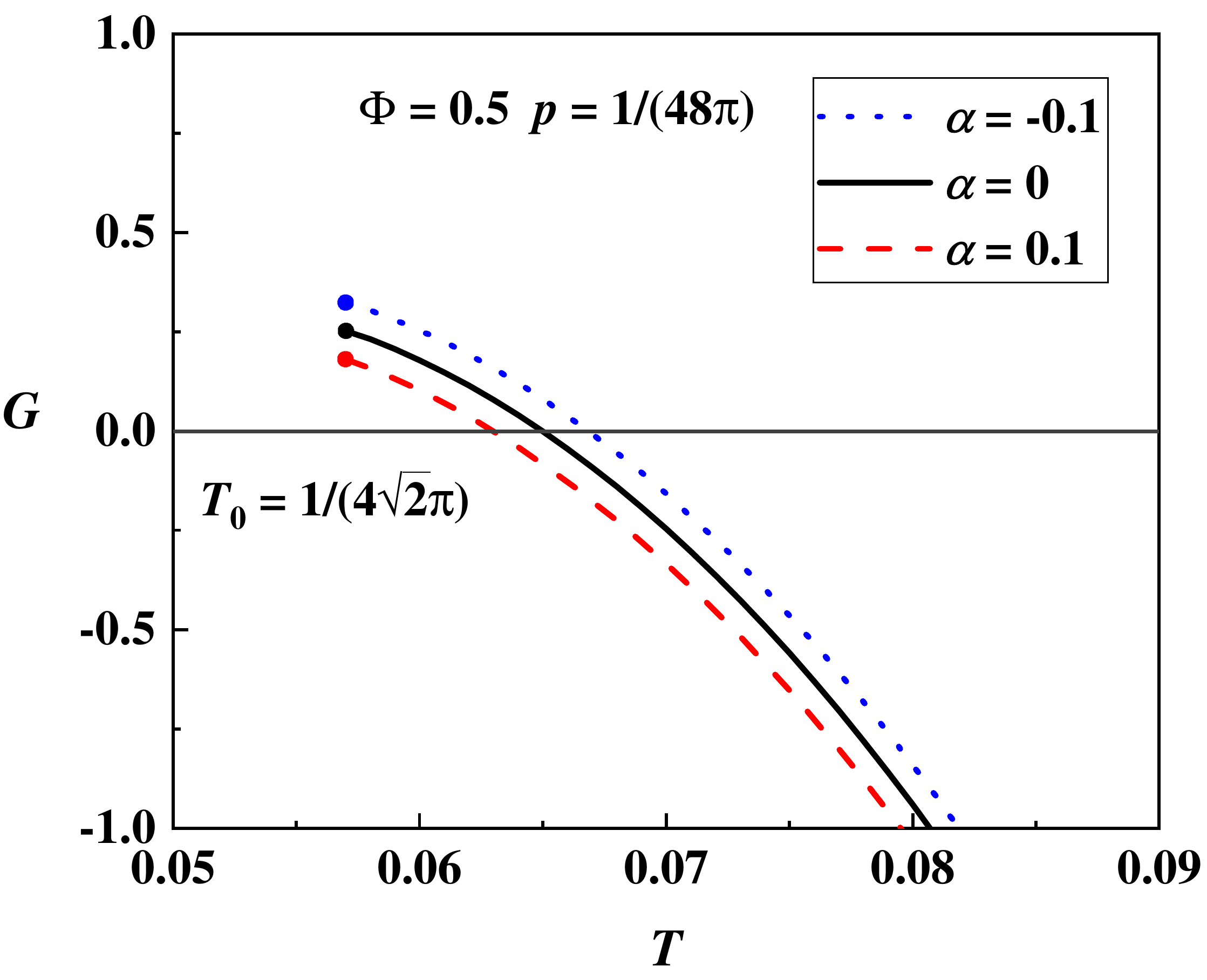}
\end{center}
\caption{The Gibbs free energy of large RN--AdS black holes as a function of temperature, with $\Phi=0.5$, $p=1/(48\pi)$, and different values of $\alpha$. All the $G$--$T$ curves start from the same minimal temperature $T_0=1/(4\sqrt{2}\pi)$, analogous to the Schwarzschild--AdS case in Fig. \ref{f:GTSAdSwith}.} \label{f:GTRNAdSwith}
\end{figure}

The $G$--$T$ curves of the RN--AdS black holes are plotted in Fig. \ref{f:GTRNAdSwith}, with $\Phi=0.5$, $p=1/(48\pi)$, and different values of $\alpha$. We observe that $T_{\rm HP}$ decreases with $\alpha$, consistent with the analysis of the coexistence lines in Fig. \ref{f:TpRNAdSwith}. Moreover, all the $G$--$T$ curves set out from the same minimal temperature $T_0$, as indicated in Eq. (\ref{t0}).

Till now, we have investigated in detail the HP phase transitions of the RN--AdS black holes with the GB term in the grand canonical ensemble with fixed electric potential. In general, there are clear similarities between the Schwarzschild--AdS and RN--AdS black holes, and these can be seen in the coexistence lines in Figs. \ref{f:TpSAdSwith} and \ref{f:TpRNAdSwith} and the $G$--$T$ curves in Figs. \ref{f:GTSAdSwith} and \ref{f:GTRNAdSwith}.

\subsection{HP phase transitions of the KN--AdS black holes} \label{sec:3}

Finally, we present the complete picture of the HP phase transitions of the most general KN--AdS black holes in the GB gravity. Because rotation breaks the spherical symmetry of horizon topology, the calculational difficulties greatly increase. Therefore, we jump the step for the rotating Kerr--AdS black holes and proceed directly to the charged and rotating KN--AdS black holes, as their complexities are almost the same. In fact, most of the calculations deal with the algebraic equations of $S$ of degrees higher than four that cannot be solved analytically, so all the results below are obtained numerically, that is, everything will be illustrated diagrammatically. We work in the grand canonical ensemble with fixed electric potential and angular velocity and concentrate on the $G$--$T$ curves of large KN--AdS black holes for simplicity.

\begin{figure}[h]
\begin{center}
\subfigure[]{\includegraphics[width=0.95\linewidth,angle=0]{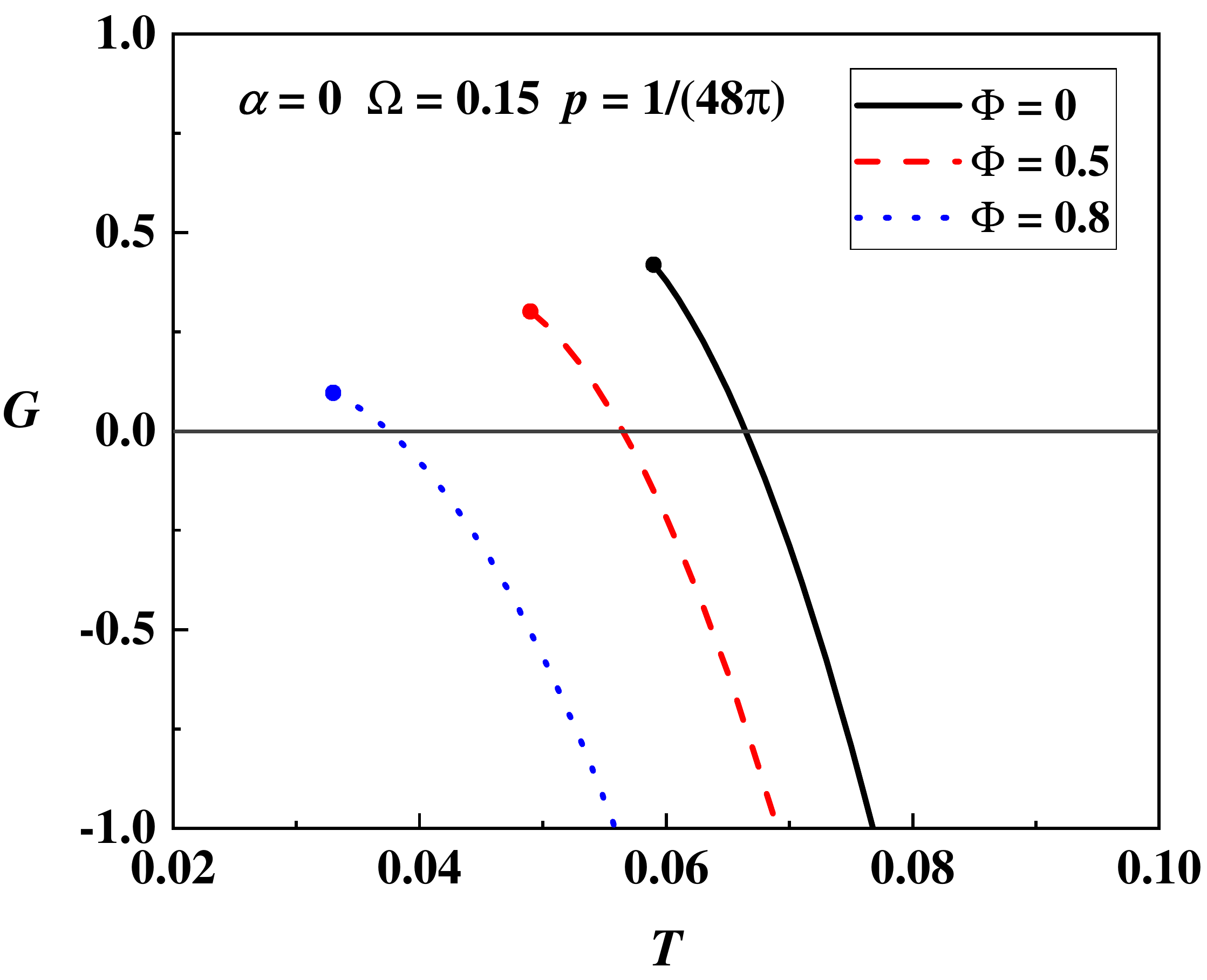} \label{Fig.sub.3}} \quad
\subfigure[]{\includegraphics[width=0.95\linewidth,angle=0]{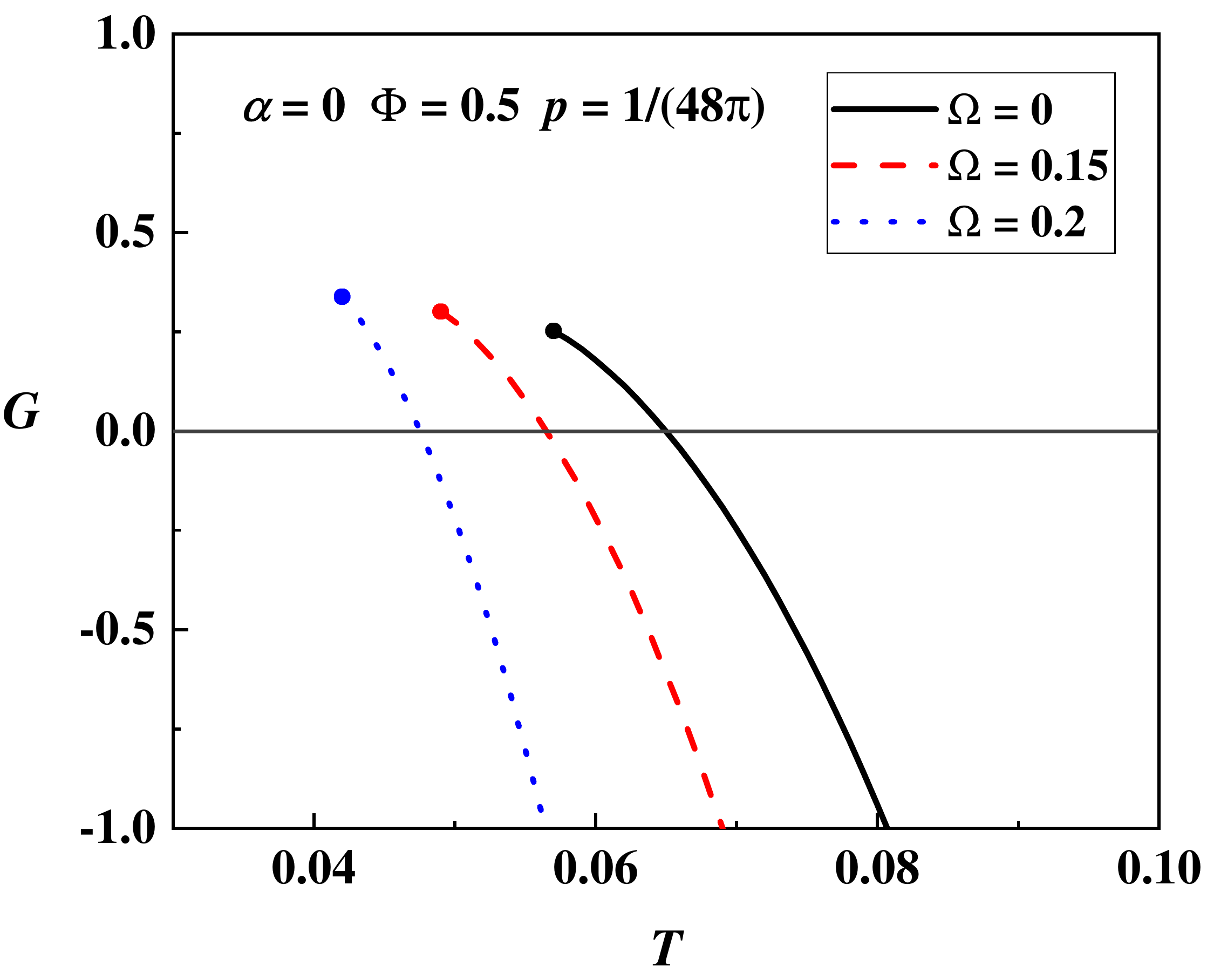} \label{Fig.sub.4}}
\end{center}
\caption{The Gibbs free energy of large KN--AdS black holes as a function of temperature, with pressure $p=1/(48\pi)$. $T_{\rm HP}$ decreases with both $\Phi$ and $\Omega$, with the detailed values of $\Phi$ and $\Omega$ listed in each panel. The KN--AdS black holes also have the minimal temperatures $T_0$, and $G(T_0)$ decreases with $\Phi$ but increases with $\Omega$.} \label{f:GTKNAdSwithout}
\end{figure}

Without the GB term, all the relevant physical variables are listed in Eqs. (\ref{M}), (\ref{T}), (\ref{Phi}), and (\ref{Omega}). By the same procedure as before, we obtain the Gibbs free energy of the KN--AdS black holes, and the $G$--$T$ curves are plotted in Figs. \ref{Fig.sub.3} and \ref{Fig.sub.4}, with different values of $\Phi$ and $\Omega$. We find that $T_{\rm HP}$ decreases with both $\Phi$ and $\Omega$ at a given pressure. This feature is qualitatively similar to that of the RN--AdS black holes.

\begin{figure}[h]
\begin{center}
\includegraphics[width=0.95\linewidth,angle=0]{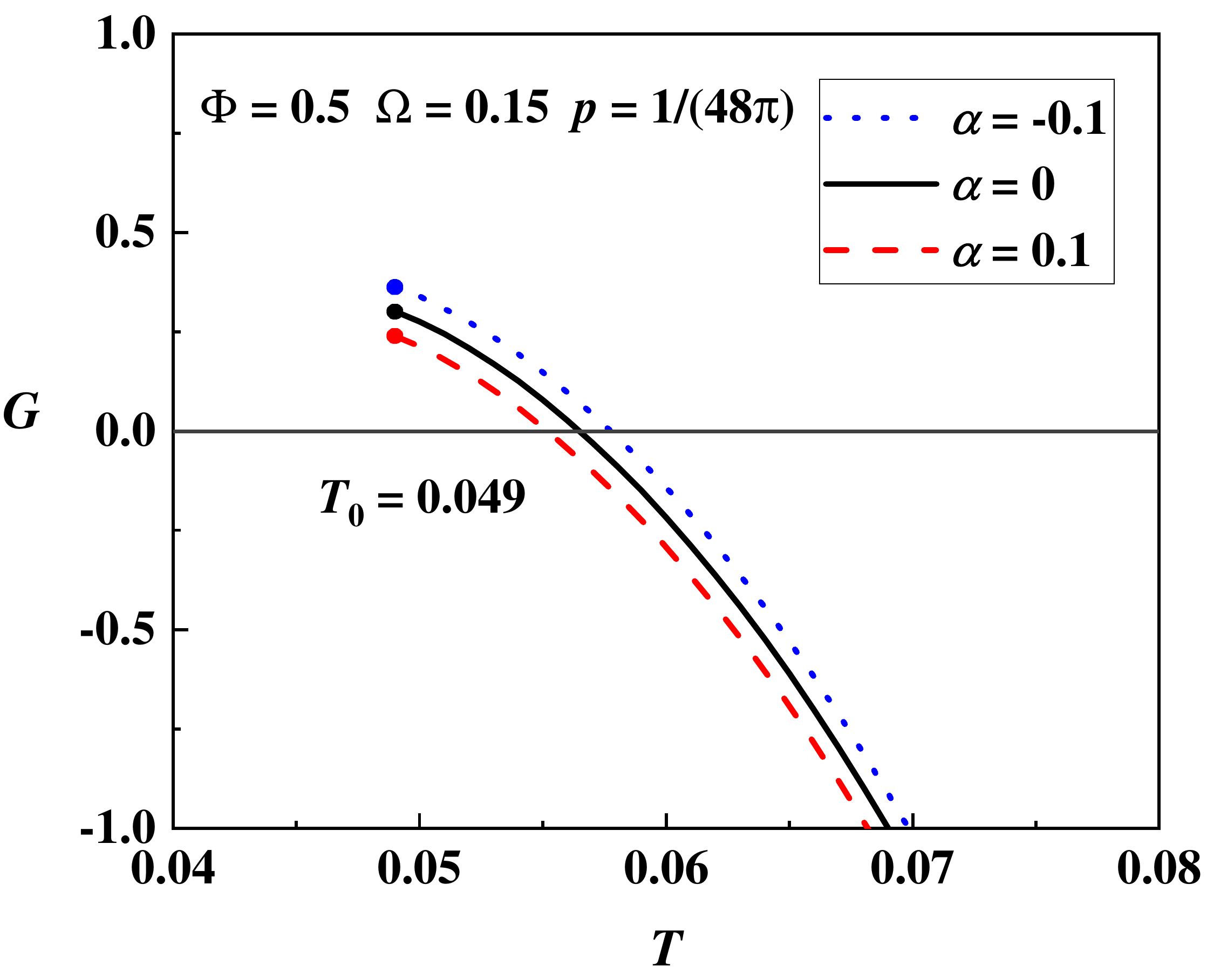}
\end{center}
\caption{The $G$--$T$ curves of the KN--AdS black holes with the GB term, with $\Phi=0.5$, $\Omega=0.15$, and $p=1/(48\pi)$. $T_{\rm HP}$ decreases with $\alpha$, and all the curves start from the same minimal temperature $T_0=0.049$. These behaviors are very similar with those of the Schwarzschild--AdS and RN--AdS black holes with the GB term.}\label{f:total}
\end{figure}

Adding the GB term makes the situation more complicated, and the $G$--$T$ curves are plotted in Fig. \ref{f:total}, with $\Phi=0.5$, $\Omega=0.15$, and different values of $\alpha$. We observe that $T_{\rm HP}$ again decreases with $\alpha$, with $\Phi$ and $\Omega$ fixed. There is still a minimal temperature $T_0$ for the KN--AdS black holes with the GB term. As a result, there exist both lower and upper bounds of $\alpha$, if the HP phase transition happens. Also, all the $G$--$T$ curves set off from the same minimal temperature $T_0$. In summary, the characteristics of the KN--AdS black holes with the GB term are still analogous to those of the Schwarzschild--AdS and RN--AdS black holes.

Here, we should make an important comment on our numerical technology. In the calculations, we first express $M$, $T$, $\Phi$, and $\Omega$ in terms of $S$ and then utilize $S$ as the intermediate variable in calculating $T_{\rm HP}$ and $G$, instead of the horizon radius $r_+$. This difference is not evident for the Schwarzschild--AdS and RN--AdS black holes with spherical horizons, but is distinct for the KN--AdS black holes with non-spherical ones. In the latter case, if we express physical quantities in terms of $r_+$, we have to use the variables $q$ and $a$. Unfortunately, they are again related to $\Xi=1-a^2/l^2$ and cannot be regarded as independent variables. Therefore, the calculations by virtue of $r_+$, $q$, and $a$ in the KN--AdS case are not only inconvenient, but also very dangerous to lead to false conclusions (e.g., without the GB term, the HP phase transition could happen only below a critical pressure, but actually they can happen at all pressures.)

\section{Conclusion} \label{sec:con}

The GB gravity is the minimal extension of the Einstein gravity, including the latter as the low energy and small curvature limit. In four-dimensions, the GB term is a topological invariant and is thus trivial to gravitational dynamics. However, it influences black hole thermodynamics via introducing the corrections to the black hole entropy and Gibbs free energy as $S+4\pi\alpha$ and $G-4\pi\alpha T$. Therefore, the GB term significantly affects one of the most important issues in black hole thermodynamics---the HP phase transition between a stable black hole and the thermal AdS space.

In this paper, the HP phase transitions of the Schwarz-schild--AdS, RN--AdS, and KN--AdS black holes in the extended phase space are systematically investigated in the GB gravity. In the extended phase space, the cosmological constant in the AdS space is effectively interpreted as a varying thermodynamic pressure $p$. Then, the HP phase transition temperature $T_{\rm HP}$ as a function of $p$ and the Gibbs free energy $G$ as a function of $T$ are calculated in detail. For the charged and rotating black holes, we work in the grand ensemble with fixed electric potential $\Phi$ and angular velocity $\Omega$. The basic conclusions of our work can be drawn as follows.

(1) The HP phase transition temperature $T_{\rm HP}$ is an increasing function of $p$. Below or above $T_{\rm HP}$, the thermal AdS or large black hole phase is thermodynamically preferred, meaning that the thermal AdS phase is more like a solid rather than a gas in the extended phase space. For charged and rotating black holes, both $\Phi$ and $\Omega$ decrease $T_{\rm HP}$. When the GB term is taken into account, the GB coupling constant $\alpha$ also decreases $T_{\rm HP}$, because it induces a correction $-4\pi\alpha T$ in the Gibbs free energy, and the $G$--$T$ curves thus move downward and intersect the $T$-axis at lower temperatures.

(2) For the black hole temperature, in the Schwarzschild--AdS case, there is a positive minimum $T_0=\sqrt{2p/\pi}$, and in the RN--AdS and KN--AdS cases, the minimal black hole temperatures still exist and are modified by $\Phi$ and $\Omega$. $T_0$ is determined by the meeting point of the two branches of the $S$--$T$ curves of the stable large black holes and the unstable small ones, and it is unchanged in the GB gravity.

(3) If the HP phase transition happens, the two corrections from the GB term, $S+4\pi\alpha$ and $G-4\pi\alpha T$, give rise to both lower and upper bounds of $\alpha$ and the corresponding upper bounds of $p$. Hence, there are terminal points in the coexistence lines. If $\alpha$ is beyond these bounds, either the black hole entropy becomes negative, or there appears some other phase behavior instead of the HP phase transition.

In summary, the physical properties of the HP phase transitions of the Schwarzschild--AdS, RN--AdS, and KN--AdS black holes in the GB gravity are qualitatively analogous. Generally speaking, electric potential, angular velocity, and the GB term reduce the HP phase transition temperature, and if the HP phase transition happens, the GB coupling constant $\alpha$ will have both lower and upper bounds. Altogether, we hope to present a whole picture of the HP phase transitions in the extended phase space in the GB gravity. As we have seen, even in the simplest Schwarzschild--AdS case, there is still some interesting issue to be explored.

We are very grateful to Zhao-Hui Chen, Peng Wang, Zhi-Zhen Wang, Shao-Wen Wei, and Ze-Wei Zhao for fruitful discussions, and also deeply thank the referee for the very valuable suggestions. This work is supported by the Fundamental Research Funds for the Central Universities of China (Nos. N170504015 and N182410008-1).

\end{document}